%% file: Sauron1.tex
\documentclass[useAMS,usenatbib,usegraphicx]{mn2e}
\usepackage{amssymb}
\usepackage{subfigure}
\usepackage{subfloat}
\newcommand{\OIII}{\hbox{[O\,{\sc iii}]}}
\newcommand{\OIIIwa}{\hbox{[O\,{\sc iii}]$\lambda $4959}}
\newcommand{\OIIIwb}{\hbox{[O\,{\sc iii}]$\lambda $5007}}
\newcommand{\OIIIww}{\hbox{[O\,{\sc iii}]$\lambda\lambda $4959,5007}}

\newcommand{\NIIwb}{\hbox{[N\,{\sc ii}]$\lambda $6583}}

\newcommand{\NIww}{\hbox{[N\,{\sc i}]$\lambda\lambda $5198,5200}}
\newcommand{\NIwa}{\hbox{[N\,{\sc i}]$\lambda $5198}}
\newcommand{\NIwb}{\hbox{[N\,{\sc i}]$\lambda $5200}}
\newcommand{\NI}{\hbox{[N\,{\sc i}]}}

\newcommand{\Hbw}{\hbox{H$\beta$\,$\lambda $4861}}
\newcommand{\Hb}{\hbox{H$\beta$}}

\newcommand{\Ha}{\hbox{H$\alpha$}}
\newcommand{\Mgb}{Mg\textit{b}}
\newcommand{\Sauron}{\hbox{\tt SAURON}}
\newcommand{\Edd}{\hbox{\rm Edd}}
\newcommand{\nodata}{ ~$\cdots$~ }
\title[Comparison of Stellar and Gaseous Kinematics of Nearby Seyfert and Inactive Galaxies.]{The Central Kiloparsec of Seyfert and Inactive Host Galaxies: a Comparison of Two-Dimensional Stellar and Gaseous Kinematics.}
\author[G. Dumas et al.]{Ga\"elle Dumas$^{1,2}$\thanks{E-mail: gdumas:obs.univ-lyon1.fr (GD)},  Carole G. Mundell$^{2}$, Eric Emsellem$^{1}$, Neil M. Nagar$^{3}$\\
$^{1}$Universit\'e de Lyon 1, CRAL, Observatoire
de Lyon, 9 av. Charles Andr\'e, F-69230 Saint-Genis Laval;\\
CNRS, UMR 5574 ; ENS de Lyon, France.\\
$^{2}$Astrophysics Research Institute, Liverpool John Moores University, Twelve Quays House, Egerton Wharf, Birkenhead CH41 1LD, UK.\\
$^{3}$Astronomy Group, Universidad de Concepci\'on, Concepci\'on, Chile.}

\begin{document}

\date{Accepted ... Received ... in original ...}

\pagerange{\pageref{firstpage}--\pageref{lastpage}} \pubyear{...}

\maketitle

\label{firstpage}

\begin{abstract}
We investigate the properties of the two-dimensional distribution and
 kinematics of ionised gas and stars in the central kiloparsecs of a
 matched sample of nearby active (Seyfert) and inactive galaxies, using
 the \Sauron\ Integral Field Unit on the William Herschel Telescope.
 The ionised gas distributions show a range of low excitation regions
 such as star formation rings in Seyferts and inactive galaxies, and high
 excitation regions related to photoionisation by the AGN.  The stellar
 kinematics of all galaxies in the sample show regular rotation patterns typical of disc-like systems, with kinematic axes which are well aligned
 with those derived from the outer photometry and which provide a reliable
 representation of the galactic line of nodes. After removal
 of the non-gravitational components due to e.g. AGN-driven outflows, the
 ionised gas kinematics in both the Seyfert and inactive galaxies are
 also dominated by rotation with global alignment between stars and gas
 in most galaxies. This result is consistent with previous findings from
 photometric studies that the large-scale light distribution of Seyfert
 hosts are similar to inactive hosts. However, fully exploiting the
 two-dimensional nature of our spectroscopic data, deviations from
 axisymmetric  rotation in the gaseous velocity
 fields are identified that suggest the gaseous kinematics are more
 disturbed at small radii in the Seyfert galaxies compared with the
 inactive galaxies, providing a tentative link between nuclear
 gaseous streaming and nuclear activity.
\end{abstract}

\input{intro}

\input{obs_reduc}

\input{results}

\input{global_kinematic}

\input{discussion_conclusion}

\bibliographystyle{mn2e}
\bibliography{these_newbib}

\input{appendices}

\label{lastpage}

\end{document}

%% file: intro.tex
\section{Introduction}

It is now widely accepted that central supermassive black holes (SMBH) are ubiquitous in bulge-dominated galaxies \citep{ferrarese_2000, Gebhardt_2000}, but ongoing nuclear activity is observed only in 10 to 20 per cent of nearby galaxies, 3 per cent at the Seyfert level \citep{kewley_2006} suggesting black hole re-ignition via accretion of material in the central parts of the galaxy is required. A key unanswered question is whether the ignition mechanism is related to the galaxy host properties in particular the fuel transportation mechanism.  Spiral galaxies contain a rich supply of potential fuel and the search for a mechanism to remove angular momentum and drive material towards the nucleus has been the focus of many studies. Numerical simulations predict efficient gas radial inflows due to non-axisymmetric perturbations in the gravitational potential such as bars or driven by galaxy interactions \citep{fueling_simu}, which may help to bring gas close to the galactic nucleus. However, most of past imaging studies failed to find any significant differences between galaxies hosting an Active Galactic Nucleus (AGN) or not, on spatial scales that encompass nearby companions/galactic interactions \citep{derobertis_1998,schmitt_2001,ho6}, stellar bars or nuclear spirals \citep{mulchaey_1997,laurikainen_2004,2003_martiniII}. In contrast, \citet{Hunt04} conducted a NICMOS/HST imaging study of the circum-nuclear regions of 250 nearby galaxies and found a significant excess of isophotal twists in the circumnuclear regions of Seyfert~2 galaxies, compared to other categories (HII/Starburst, Seyfert~1, LINER, inactive galaxies). This result may hint at an evolutionary scenario to explain the different states in which a nearby galaxy is observed, and could suggest the presence of identifiable dynamical differences between Seyfert and inactive galaxies in the central kpc regions. 

Probing the dynamics of Seyfert galaxies requires spectroscopic data. Single aperture and long-slit spectroscopy studies are clearly inadequate to investigate the complex structures observed in the central kpc of Seyfert galaxies. Two-dimensional spectroscopy (integral-field spectroscopy, IFS hereafter) is therefore a pre-requisite to study the dynamics of the gaseous and stellar components. Early work with such instruments provided two-dimensional gas and stellar kinematic maps of the central parts of a few nearby active galaxies, in optical wavelength: e.g. NGC\,4151 \citep{4151_mediavilla95}, NGC\,3227 \citep{3227_nature93}, NGC\,1068 \citep{1068_garcia97} with INTEGRAL (an IFS based on optical fibers), or NGC\,2110 \citep{NGC2110_1} using OASIS (an IFS using a microlens array), as well as in the near infrared (NIR): The Circinus galaxy observed with the NACO spectrograph and SINFONI at the VLT \citep{prieto_2004, circunus_2006}, NGC\,3227 \citep{3227_1} observed with SINFONI. Multiple gaseous systems and kinematic perturbations are sometimes revealed, but the corresponding field-of-view (FOV) and/or the angular resolution were often too small to disentangle the AGN-related and galactic disc line emissions. Statistical conclusions cannot be drawn from these detailed studies of individual Seyfert galaxies, originally targeted for their complex nuclear properties and lacking any control inactive galaxy comparison. More recenly, the molecular gas in the central part of a larger (though still small) sample of Seyfert galaxies was observed using the Plateau de Bure Interferometer in the course of the NUGA survey \citep{NUGA_I}, revealing the potential role of gravity torques in the feeding of the inner 100~pc \citep[e.g.][]{nugaIV}. Gas reponds non-linearly to deviations from axisymmetry so a more direct link with the gravitational potential is still difficult to establish, due to the lack of large-scale stellar kinematic maps for these galaxies. Six nearby Seyferts were also observed with GMOS/Gemini, enabling the mapping of the ionised gas and stellar kinematics in the central few arcseconds \citep{barbosa_2006}. IFS such as SINFONI/VLT reached regions even closer to the nucleus \citep{3227_1}, thanks to the use of adaptive optics. Again, such observations are useful for studying the kinematics in the nuclear regions, identifying the putative molecular torus around the SMBH, retrieving the characteristics of the central engine such as the mass of the SMBH, or studying the environment of the active nucleus. However, observed kinematics at these relatively small scales are rather difficult to interpret in the context of the host galaxy, considering the significant influence of non-gravitationally driven processes (e.g. outflows). 

The \Sauron\  IFS, mounted on the William Herschel Telescope (La Palma, Spain) has  FOV ($33\arcsec\times41\arcsec$ in low resolution mode) that is large enough to observe ionised gas under the influence of the galaxy host potential, and a high enough spatial sampling to still probe the AGN-related emission in the inner parts. These data are then well suited to study both the stellar and gaseous kinematics in the above-mentioned context of fueling. A number of key studies were recently performed with \Sauron\, providing unique datasets for a sample of nearby early-type \citep{paper2,paper3, paper5} and spiral galaxies \citep{ganda_2006, paper7}. As a spin off, the distribution and kinematics of the gas and stellar components were also obtained and studied for some well known Seyfert (NGC\,1068, \citealp{1068_1}, NGC\,5448, \citealp{Fathi_2005}) or spiral inactive galaxies like M\,100 \citep{M100_2006}. These works provided evidence for the existence of ionised gaseous inward streaming in the inner few kpc, hinting at a mechanism for transporting gas in the circumnuclear regions and further. However, no survey has been yet pursued to map the two-dimensional kinematic and morphology in the central parts of a well-selected sample of Seyfert and inactive galaxies, in order to search for and quantify potential dynamical differences between these two populations at these spatial scales.

We therefore designed a new IFS survey for a well-defined sample of active and control inactive galaxies, selected from the RSA sample with nuclear classification of \citet{Ho3}.
This program aims at probing the galactic potential of the host galaxies by comparing the properties of both the ionised gas and stellar components in the circumnuclear regions of Seyfert galaxies. We observed these galaxies with the \Sauron\  IFS, which provides a spatial coverage sufficient  to map regions from the central $33\arcsec\times41\arcsec\sim$ few kpc down to the inner $1\arcsec\sim 50$~pc at 10~Mpc. In Sect.~\ref{sec:2}, we present the sample selection, the observations and data reduction. In Sect.~\ref{sec:result_intro}, we present the associated stellar and ionised gas maps. A kinematic analysis is presented in Sect.~\ref{sec:ana}. We then briefly discuss our results in a more general context and conclude in Sect.~\ref{sec:discussion_conclusion}.

%% file: obs_reduc.tex
\section{Observations and Analysis}
\label{sec:2}
\subsection{Sample Selection}
\label{sec:sample}
\begin{figure*}
\includegraphics[width=\textwidth]{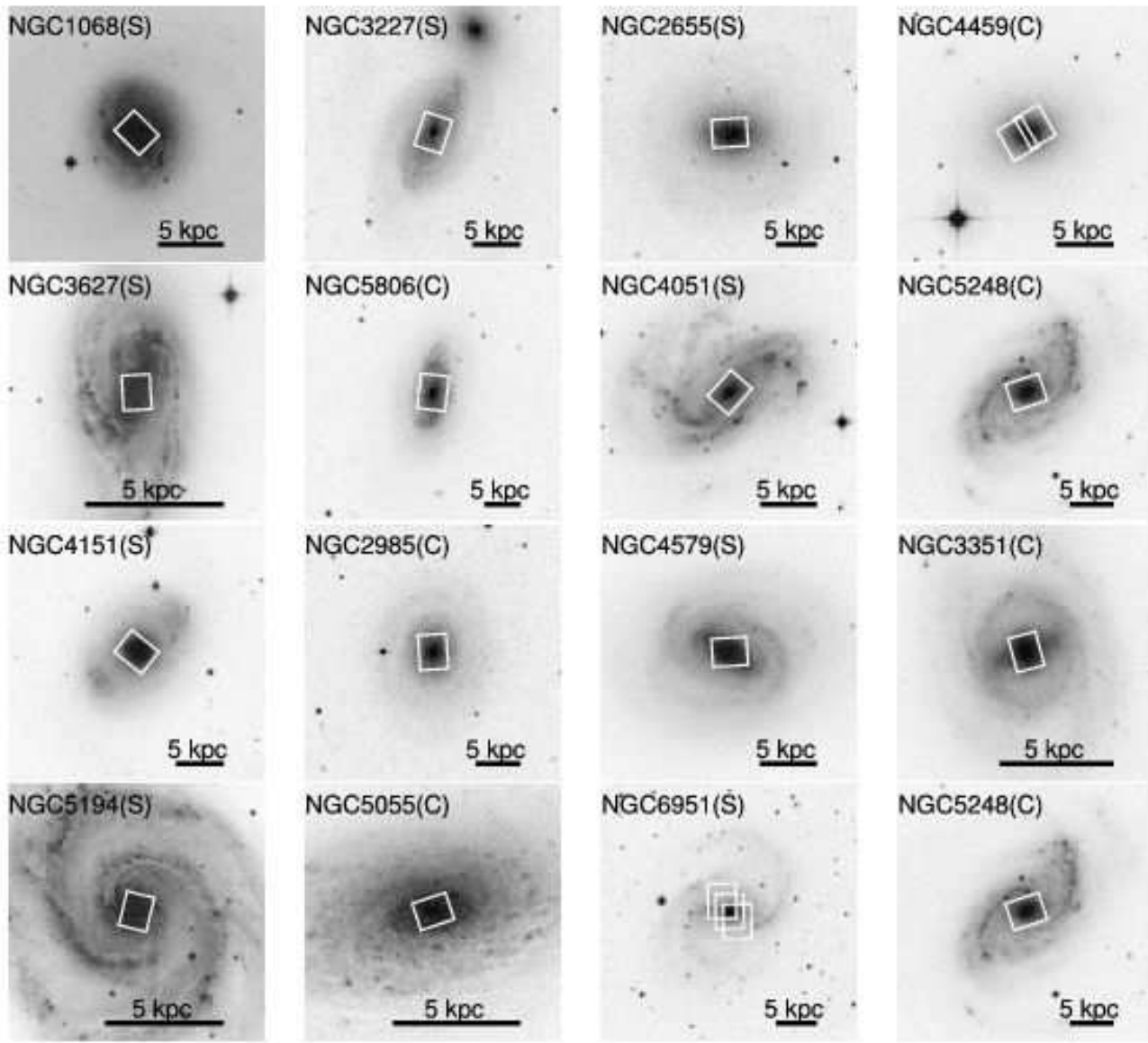}
\caption{R-band Digital Sky Survey images of the sample galaxies. The '(S)' or '(C)' on the right of the object names stand for Seyfert or Control galaxy, respectively. Each Seyfert galaxy is displayed on the left of its associated control galaxy, except for NGC\,1068 and NGC\,3227 (first two panels) for which no control has been observed. The orientation is such that North is up and East is left, the bar located at the bottom right corner of each panel corresponds to the spatial length of 5 kpc. Overplotted on each image is the position of the \Sauron\  field of view for that galaxy.}
\label{dss}
\end{figure*}

\begin{table*}

\centering
% \begin{minipage}{140mm}
\begin{tabular}{|l|l|l|c|r|r|c|r|r|c|c|c}
\hline
Pairs & Name&Type& T&Distance&$D_{25}$& Inclination& $B_T$&Vsys&$M_V$(nuclear) & Spectral  & Ref. \\
%Pairs & Name&Type\footnote{NED}& T\footnote{LEDA}&Distance\footnote{\citet{Ho3}}&$D_{25}$\mpfootnotemark[\value{mpfootnote}]& Inclination& $B_T$\mpfootnotemark[\value{mpfootnote}]&Vsys$^{a}$&$M_V$(nuclear)\mpfootnotemark[\value{mpfootnote}] & Spectral Classification\mpfootnotemark[\value{mpfootnote}] & Ref. \\
	&NGC	&		&	&(Mpc)	&(arcmin)	&(deg)	&(mag)	&(km\,s$^{-1}$)	&(mag)		&Classification	&\\
(1)	&(2)	&(3)		&(4)	&(5)	&(6)		&(7)	&(8)	&(9)		&(10)		&(11)	&(12)\\
\hline
	& 1068	&SA(rs)b	&3	&14.4	&7.08		&29	&9.61	&1135		&$-$20.0		&S1.8	&1\\
\cline{2-12}
	& 3227	&SAB(s)a pec	&1.2	&20.6	&5.37		&56	&11.10	&1146		&$-$19.8		&S1.5	&2 \\
\hline
1	& 2655	&SAB(s)0/a	&0.1	&24.4	&4.90		&34	&10.96	&1403		&$-$20.9		&S2	&3\\
	& 4459	&SA(r)0		&-1.4	&16.8	&3.55		&41	&11.32	&1202		&		&T2	&3\\
\hline
2	& 3627	&SAB(s)b	&3.0	&6.6	&9.12		&65	&9.65	&727		&$-$17.2		&S2/T2	&3\\
	& 5806	&SAB(s)b	&3.3	&28.5	&3.09		&58	&12.40	&1359		&		&H	&4\\
\hline
3	& 4051	&SAB(rs)bc	&4.2	&17.0	&5.25		&43	&10.83	&720		&$-$18.2		&S1.2	&3\\
	& 5248	&SAB(rs)bc	&4.2	&22.7	&6.17		&40	&10.97	&1152		&		&H	&5\\
\hline
4	& 4151	&SAB(rs)ab	&2	&20.3	&6.31		&21	&11.50	&992		&$-$19.7		&S1.5	&6\\
	& 2985	&SA(rs)ab	&2	&22.4	&4.57		&38	&11.18	&1218		&		&T2	&3\\
\hline
5	& 4579	&SAB(rs)b	&2.8	&16.8	&5.89		&38	&10.48	&1521		&$-$19.4		&S1.9/L1.9	&3\\
	& 3351	&SB(r)b		&2.9	&8.1	&7.41		&56	&10.53	&778		&		&H	&7\\
\hline
6	& 5194	&SA(rs)bc pec	&4.2	&7.7	&11.22		&20	&8.96	&463		&$-$16.0		&S2	&8\\
	& 5055	&SA(rs)bc	&4	&7.2	&12.59		&56	&9.31	&504		&		&T2	&3\\
\hline
7	& 6951	&SAB(rs)bc	&3.9	&24.1	&3.89		&34	&11.64	&1424		&$-$17.9		&S2	&9\\
	& 5248	&SAB(rs)bc	&4.2	&22.7	&6.17		&40	&10.97	&1152		&		&H	&5\\
\hline
\end{tabular}
\caption{Properties of our sample. (1) Pairs identifier, (2) Galaxy name, (3) Hubble Type (NED), (4) Numerical morphological type (LEDA), (5) Distance in Mpc \citep{Ho3}, (6) $D_{25}$ in arcmin \citep{Ho3}, (7) disc inclination in degrees,  (8) Total apparent magnitude $B_T$ in mag \citep{Ho3}, (9) Systemic velocity in km\,s$^{-1}$ (NED), (10) V-band absolute magnitude of the Seyfert nuclei $M_V$ in mag \citep{Ho3}, (11) Spectral classification \citep{Ho3}: H=HII nucleus, S=Seyfert nucleus, L=LINER, T=transition object, 'T2' implies that no broad \Ha\ emission lines was detected \citep{Ho3}. (12) References for the disc inclination values. 1: \citet{garciagomez_2002}, 2: \citet{3227_carole_95b}, 3: \citet{Ho3}, 4: \citet{Kassin_2006a}, 5: \citet{5248_Jogee_02}, 6: \citet{4151_Pedlar_92}, 7: \citet{erwin_2005}, 8: \citet{tully_M51}, 9: \citet{6951_1} \label{table:properties} }
%\end{minipage}
\end{table*}

The observations presented here form part of a larger multiwavelength
campaign to investigate host galaxy structure and kinematics on size
scales ever-closer to the nucleus, aimed at identifying or eliminating
possible triggering and fueling mechanisms \citep{mundell03, nugaIV}.  Our master sample consists
of 39 Seyfert galaxies selected from the RSA catalogue paired with 39 control galaxies with carefully matched
optical properties: $B_T$, Hubble type, inclination and, where
possible, $V_{sys}$ ($\leq4000$~km\,s$^{-1}$). 3D
spectroscopic imaging with the VLA and WSRT is underway to map the
large-scale neutral hydrogen H{\sc i} distribution and kinematics for these galaxies 
\citep{mundell_07}. Current H{\sc i} imaging
interferometers however cannot routinely resolve structures smaller
than $\sim20\arcsec$ (i.e. $r<700$~pc for our sample); studying the
distribution and kinematics of ionised gas and stars with the integral-field spectrograph \Sauron\ 
provides the missing link between the large-scale H{\sc i} disc
properties and the very central nuclear regions.

%\begin{figure*}
%\includegraphics[width=17.7cm]{pair_properties.eps}
%\caption{distribution of Hubble types, disc inclinations, absolute B magnitudes and D25 for the Seyfert and control pairs.}
%\label{prop}
%\end{figure*}

For our \Sauron\  study, we selected a distance-limited sub-sample of
Seyfert$+$control galaxies with $V_{sys}<1600$~km\,s$^{-1}$ to ensure
that Fe stellar absorption lines lie well within the spectral band for
the full \Sauron\  field-of-view. Measurement of these Fe lines is
critical in the presence of emission lines from ionised gas that
contaminate Mg lines (see Sect.~\ref{sec:starsreduc}). In total, our sub-sample
comprises 15 pairs of Seyferts$+$inactive galaxies. We completed observations of 7 pairs and two well-known Seyferts (NGC\,1068
and NGC\,3227) with no control galaxy data. NGC\,1068 has been studied
in detail by \cite{1068_1} and is included here for
completeness, while weather constraints prevented observations of the
control galaxy for NGC\,3227. Table \ref{table:properties} lists the properties of the
sample galaxies. The V-band absolute
magnitudes of the Seyfert nuclei in this sub-sample span the full
magnitude range of the 25 brightest Seyfert nuclei in the RSA
catalogue ($-20.9<M_V<-16.0$), thus offering a representative
selection of Seyfert activity.

\subsection{\Sauron\ Observation and Data Reduction}
\label{sec:obs}
\begin{table}

\centering
\begin{tabular}[h]{|l|l||l|l|c|}
\hline
Pairs	&Name	&dates		&$T_{exp}\ $	&FWHM$_{PSF}$ \\
	&NGC	&		&(s)		&(arcsec)\\
(1)	&(2)	&(3)		&(4)		&(5)\\
\hline
	&1068	&2002 January	&300 + 3x1800	&1.9$\pm$0.1\\
\cline{2-5}
	&3227	&2004 March	&3x1800		&1.4$\pm$0.2\\
\hline
1	&2655	&2004 March	&6x1800		&1.4$\pm$0.1\\
	&4459	&2001 March	&4x1800		&1.6$\pm$0.1\\
\hline
2	&3627	&2004 March	&4x1800		&1.6$\pm$0.1\\
	&5806	&2004 March	&2x1800		&2.1$\pm$0.1\\
\hline
3	&4051	&2004  March	&6x1800		&1.4$\pm$0.2\\
	&5248	&2004 March	&5x1800		&1.9$\pm$0.1\\
\hline
4	&4151	&2004 March	&3x1800		&2.0$\pm$0.3\\
	&2985	&2004 March	&3x1800		&1.2$\pm$0.1\\
\hline
5	&4579	&2004 March	&4x1800		&1.9$\pm$0.1\\
	&3351	&2004 March	&4x1800		&0.9$\pm$0.1\\
\hline
6	&5194	&2004 March	&4x1800		&1.2$\pm$0.1\\
	&5055	&2004 March	&6x1800		&2.4$\pm$0.1\\
\hline
7	&6951	&2003 August	&3x1800	&1.4$\pm$0.1\\
	&5248	&2004 March	&5x1800		& 1.9$\pm$0.1\\
\hline
\end{tabular}
\caption{Details of the exposures of our sample. (1) Pairs identifier, (2) NGC\, number, (3) date of observation  (4) Exposure time in sec, (5) Seeing, Full Width at Half Maximum in arcseconds. \label{table:observations}}
\end{table}

Observations of our sample were carried out between 2001 and 2004 using the integral-field spectrograph \Sauron\  at the 4.2m William Herschel Telescope (WHT) at La Palma, Spain. Table \ref{table:observations} summarizes the exposure time for each pointing and lists the equivalent seeing of final reduced data.

The low spatial resolution mode of \Sauron\  was used, providing a field-of-view (FOV) of  $33\arcsec\times41\arcsec$ with a square sampling of 0.94 arcsec per spatial element (lens).  This delivers about 1500 spectra simultaneously per object which cover the spectral range 4825-5380 \AA, with a resolution of 4.2 \AA. This wavelength range includes a number of important stellar absorption lines (\Hb, Fe5015, \Mgb, Fe5270) and ionised gas emission lines (\Hb, \OIII\ and \NI). More details on the \Sauron\  spectrograph can be found in~\cite{paper1}.  Fig. \ref{dss} presents the \Sauron\  field-of-view overlaid on R-band Digital Sky Survey (DSS) images of our galaxies.

The data of the 15 galaxies were reduced using the dedicated X\Sauron\  software and an automatic pipeline available within the \Sauron\  consortium \citep{paper1,paper2}. The main steps include: bias and dark subtraction, extraction of the spectra using a mask, wavelength calibration, low-frequency flatfielding, cosmic rays removal, homogenisation of the spectral resolution in the field, sky subtraction (using 146 sky spectra 1.9 arcmin away from the main field). Then the flux calibration was applied as explained in detail in \cite{paper6}. 
The flux calibrated individual exposures are then accurately centred with respect to each others and merged. The final merged datacube is sampled onto a squared grid with $0\farcs8 \times0\farcs8$ pixels.

Finally, the point spread function (PSF) of each merged exposure was determined by comparing the reconstructed \Sauron\  intensity distribution with HST/WFPC2 images. The \Sauron\  PSF is modeled by a single 2D Gaussian, whose parameters is determined by minimizing the differences between the HST/WFPC2 images, convolved by this Gaussian and the \Sauron\  reconstructed images. The derived seeing values (Full Width at Half Maximum) are listed in Table \ref{table:observations}.

\subsection{Derivation of Stellar Kinematics}
\label{sec:starsreduc}

In order to ensure the  reliability of the stellar kinematic measurements, the merged datacubes were spatially binned using the Voronoi 2D binning algorithm of~\cite{binning}, creating bins with a minimum signal-to-noise ratio (S/N) = 60 per bin. The stellar contribution to the \Sauron\  spectra is potentially contaminated by emission lines: \Hbw, \OIIIww\ and \NIww, therefore we first identified and masked out the spectral regions that are significantly contaminated by emission. The stellar kinematics are measured on each spectrum in our binned datacubes using the penalized pixel fitting (pPXF) method developed by~\cite{PXF}. The line-of-sight velocity distribution (LOSVD hereafter) is parameterised by  Gauss-Hermite series and the algorithm determines the mean velocity V, the velocity dispersion $\sigma$ and the higher Gauss-Hermite moments $h_3$ and $h_4$ which minimize the differences between the observed spectrum and a stellar template spectrum convolved with the corresponding LOSVD. A low-order polynomial is included to account for small differences between the galaxy and the template spectra. A first estimate of the kinematic parameters is done using a single star spectrum as template. Then using this initial estimate, an optimal stellar template is derived for each individual spectrum, via the use of a large library of stellar templates \citep{vazdekis}. We finally iterate by measuring best-fitting values at each position for the velocity V, the velocity dispersion $\sigma$, $h_3$ and $h_4$ using this time the optimal templates obtained from the previous step. 
The stellar kinematic parameters were determined using this method automatically implemented in the \Sauron\  data reduction pipeline for our whole sample except for the Seyfert 1 galaxies: NGC\,3227, NGC\,4051 and NGC\,4151. For these galaxies, the automatic pipeline alone is inadequate due to the presence of a broad \Hb\ emission line in the inner few arcseconds (see Fig. \ref{fit}.c). The spectral regions that are masked have to be carefully defined, and then we fit the stellar kinematics of these 3 galaxies interactively. 

The AGN continuum is not separated from the stellar component using this technique. The stellar kinematic derivation of the Seyfert 2s is a priori not affected by the AGN continuum since the central engine is hidden from direct view. The AGN continuum would however affect the derivation of the velocity dispersion of the Seyfert 1s in the central regions where it is dominant. These regions have been excluded from the stellar kinematic analysis and subsequent discussion.

\subsection{Derivation of Gaseous Distribution and Kinematics}
\label{sec:gasreduc}

\begin{figure*}
\includegraphics[width=\textwidth]{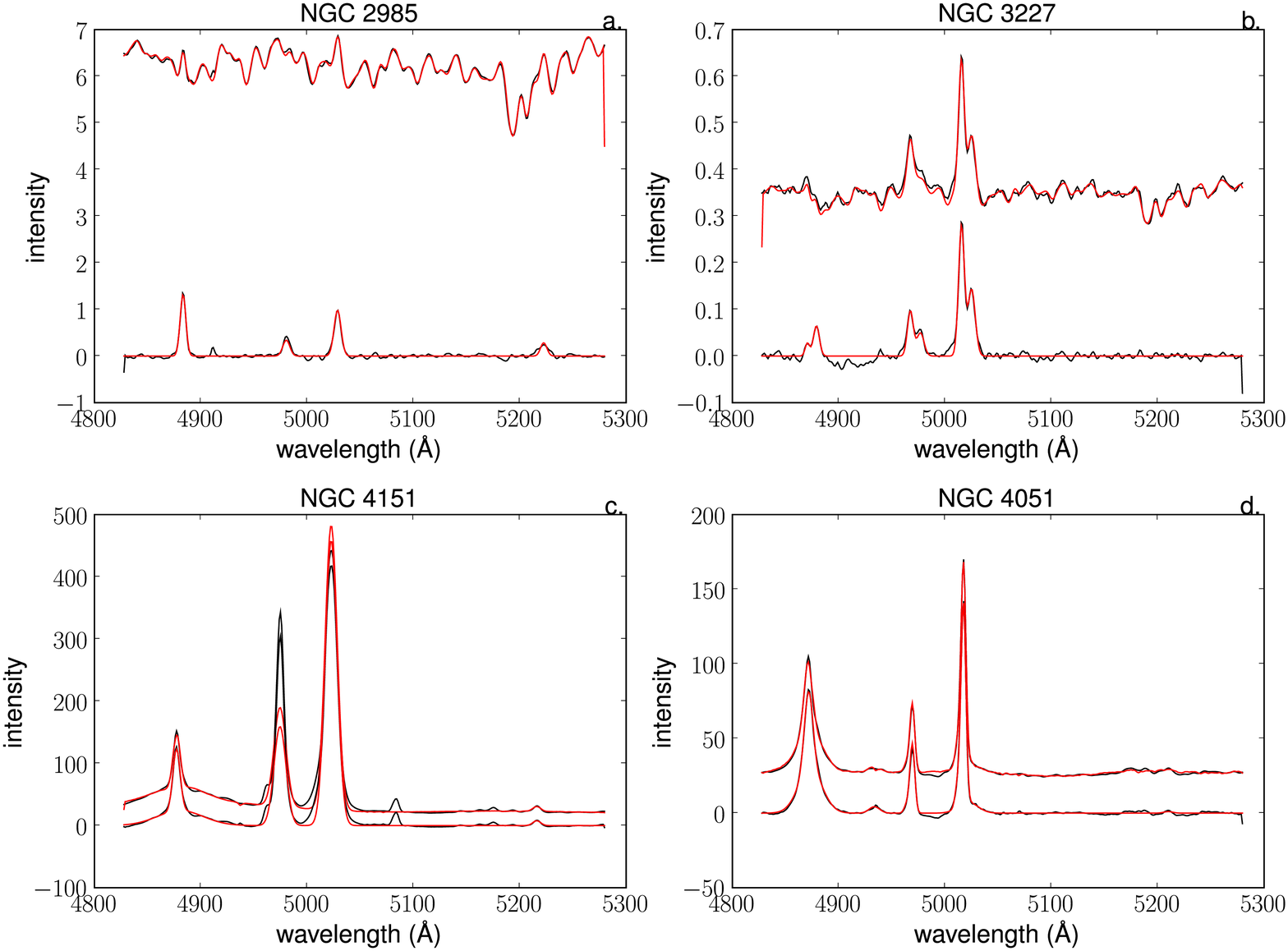}
\caption{Examples of spectra and their corresponding fits for 4 galaxies. (a): NGC\,2985 (inactive), (b): NGC\,3227 (Seyfert 1.5), (c): NGC\,4151 (Seyfert 1.5), (d): NGC\,4051 (Seyfert 1.2). In each panel, the black line on the top shows the galaxy spectrum and the red line, its corresponding fit which is composed of the best-fitting stellar template added to the emission line fit. Below this spectrum, the pure emission line spectrum is shown in black. The corresponding emission lines fit is overplotted in red. The spectra were taken in the central pixel of the FOV, except for NGC\,3227, for which the spectrum was taken at 4$\arcsec$ North-East from the center.}
\label{fit}
\end{figure*}

\begin{figure*}
\includegraphics[width=\textwidth]{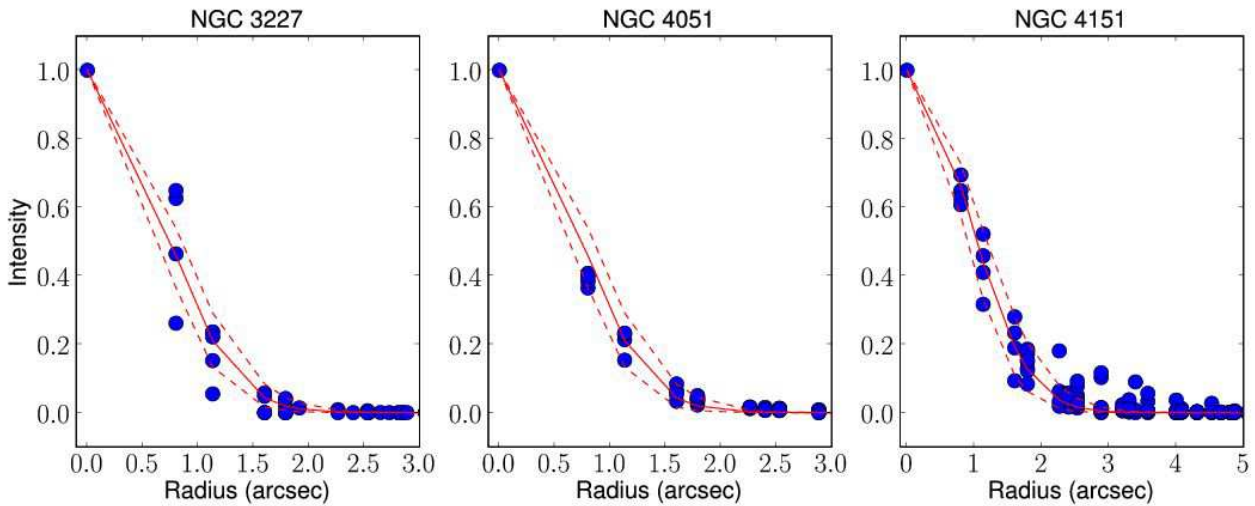}
\caption{Radial normalised profiles of the BLR for NGC\,3227, NGC\,4051 and NGC\,4151 (filled blue circles). The red solid line corresponds to the best-fit seeing value for each of the three galaxies (see Table~\ref{table:observations}), the two red dashed curves correspond to the associated upper and lower limits at 2$\sigma$.}
\label{BLRr}
\end{figure*}

The stellar continuum resulting from the stellar kinematic fit was subtracted from the original data, providing pure emission line datacubes. The wavelength range of our observations includes the \Hbw, \OIIIww\ and \NIww\ emission lines. The parameters of these emission lines (intensity, mean velocity and FWHM) were derived from Gaussian fitting  using the \hbox{\tt fit/spec} software developed by~\cite{fitspec}. 

The fit was performed on two systems of emission lines~: the first consisting of the  \Hb\ and \NI\ lines, and the second, of the \OIII\ lines. The \Hb\ and \OIII\ lines were fitted independently in order to detect differences in the kinematics of the two lines. Since the \NI\ lines are significantly weaker than the \Hb\ and \OIII\ lines, their kinematics can not be constrained independently, therefore this doublet is fitted together with the \Hb\ line system. Within each system, the lines were assumed to share the same velocity and FWHM. Constraints were applied on the parameters: the line ratio \OIIIwa/\OIIIwb\ was assumed throughout to be equal to 2.88, and $0.5 <$ \NIwa/\NIwb\ $< 1.5$. Moreover the velocity V and the velocity dispersion FWHM of the lines were bounded: V around the systemic velocity of the galaxy, and the FWHM by the spectral resolution of \Sauron\  FWHM$_{\Sauron}=4.2$ \AA. Usually the line profiles were simple enough to be fitted automatically. A first fit was done using one single Gaussian profile and this automated fit has been visually controlled  for each galaxy. An example of the emission lines fitting results is given in Fig.~\ref{fit}.a. A small number of galaxies show complex emission lines profiles and required a manual fit with additional emission line components. Additional components for \OIII\ lines, \Hb\ line or both were required in the case of five active galaxies: NGC\,3227 (Fig.~\ref{fit}.b), NGC\,4051, NGC\,4151 and NGC\,5194. These additional components are certainly associated with the Narrow Line Regions (NLR) of these galaxies. An unresolved central \Hb\ broad component has also been added in the case of the Seyfert 1 galaxies (NGC\,3227, NGC\,4051 and NGC\,4151) which corresponds to the Broad Line Regions (BLRs). The FWHM of this component is 2600~km\,s$^{-1}$, 1500~km\,s$^{-1}$ and 3100~km\,s$^{-1}$ for NGC\,3227, NGC\,4051 and NGC\,4151, respectively, and the spatial extent of the radial profile of the BLRs (Fig.~\ref{BLRr}) is consistent with the PSF derived above, within the derived uncertainties (Table~\ref{table:observations}). 
Finally, in the case of NGC\,4051, FeII blends are detected as shown in Fig.~\ref{fit}.d, and these emission lines were therefore added in the fit for this galaxy.

%% file: results.tex
%\section{Data Maps: Stellar and ionised Gas Distribution and Kinematics}
\section{Results}
\label{sec:result_intro}

In this Section, we present the distribution and kinematic maps of the stellar and ionised gas components for our sample of galaxies. Figures \ref{pair0} to \ref{pair7} present our \Sauron\  maps: stellar continuum, \OIII\ and \Hb\ intensity distributions and emission line ratio \OIII/\Hb, as well as stellar and gaseous kinematics (velocity and velocity dispersion). In the case of the Seyfert 1 galaxies, the line ratio \OIII/\Hb\ is computed using the narrow \Hb\ component. The galaxies are displayed by pair: the Seyfert on the top, its control galaxy below, except for NGC\,1068 and NGC\,3227 which are shown together (Fig.~\ref{pair0}). All the maps are oriented so that the outer photometric major-axis of the galaxy is on the horizontal axis. 
 To display the gas maps, we show only the reliable emission, i.e. when the ratio of the fitted amplitude to the surrounding noise is larger than 3. The \NI\ doublet is detected in most of our galaxies, but it is very weak, so it is not discussed further. The gaseous kinematic maps (velocity and dispersion) correspond to the \OIII\ emission lines kinematics, except for the inactive galaxy NGC\,4459 (Fig.~\ref{pair1}) for which the \Hb\ line was used since it is slightly more extended and has a better signal-to-noise ratio (Fig.~\ref{pair2}). 
 The \Hb\ emission line velocity and velocity dispersion maps  are shown in Appendix~\ref{app:HB_kin} (\OIII\ corresponding maps for NGC\,4459).

Finally, NGC\,4459 and NGC\,6951 were observed as part of other programmes \citep{paper2} with multiple fields to provide a mosaic (Fig.~\ref{dss}). For consistency with the other galaxies in our sample, we extracted images corresponding to one single \Sauron\  field of view (FOV) from the mosaiced exposures for these two galaxies. 

In the following we present the general properties of our sample. Detailed descriptions of the maps for each galaxy can be found in Appendix~\ref{app:gal_indiv}.

\subsection{Stellar and Ionised Gas Distribution}
For each galaxy, the stellar continuum maps (Figs.~\ref{pair0}-\ref{pair7}, first panel for each galaxy) were derived by integrating over the full wavelength window the spectra corresponding to the optimal stellar template obtained as explained in Sect.~\ref{sec:starsreduc}.  We constructed the ionised gas intensity maps and the \OIII/\Hb\ lines ratio maps directly from the fit of the emission lines spectra.

\subsubsection{Stellar Continuum Distribution}
Half of the galaxies in our sample present symmetric stellar continuum maps with regular isophotes. NGC\,2985, NGC\,4151 (Fig.~\ref{pair4}) and NGC\,4459 (Fig.~\ref{pair1}) show rather round central features while the isophotes are more flattened for NGC\,3227 (Fig.~\ref{pair0}), NGC\,4051 (Fig.~\ref{pair4}), NGC\,5806 (Fig.~\ref{pair2}) and NGC\,6951 (Fig.~\ref{pair7}). The latter four galaxies host a large scale stellar bar, their stellar continuum reflecting the elongation of the bar in the central regions.

Five Seyfert 2 galaxies and one inactive galaxy present more complex stellar distribution maps.  Isophotal twists or irregular isophotes are observed in five Seyfert 2 galaxies: NGC\,1068 (Fig.~\ref{pair0}), NGC\,2655 (Fig.~\ref{pair1}), NGC\,3627 (Fig.~\ref{pair2}), NGC\,4579 (Fig.~\ref{pair5}) and NGC\,5194 (Fig.~\ref{pair6}). The inactive galaxy NGC\,5055 presents an asymmetric structure (Fig.~\ref{pair6}). The surface brightness is higher on the North-West side of the FOV, the emission line flux being absorbed by the dust on the South-East (see Fig.~\ref{dss}). The predominance of irregular isophotes and twists in the circumnuclear regions of Seyfert 2 galaxies compared to Seyfert 1 or inactive galaxies has been described quantitatively by \citet{Hunt04}.

Finally, two inactive galaxies, NGC\,3351  and NGC\,5248 exhibit a circumnuclear ring at a radius of $\sim$ 5$\arcsec$ which corresponds to 550 pc for NGC\,3351 and 200 pc for NGC\,5248, respectively. These structures have been described by \citet{3351_1} and \citet{5248_1}.

\subsubsection{Ionised Gas Distribution}

Ionised gas is detected in all the galaxies of our sample, over the full \Sauron\  FOV except for NGC\,4459 (Fig.~\ref{pair1}), where emission lines are very weak outside the inner 10$\arcsec$. In this Section we describe the maps of the gas properties mainly focusing on regions which are not dominated by non-gravitational motions driven by the active nucleus.

A variety of structures can be seen in the \OIII\ and \Hb\ intensity maps:
\begin{itemize}
\item Spiral-like structures are seen in two Seyfert galaxies: NGC\,1068 (Fig.~\ref{pair0}, \Hb\ intensity map) and NGC\,4579 (Fig.~\ref{pair5}). Such structures correspond well with the nuclear molecular spirals observed by \cite{eva_1068} and \cite{nugaIV} in NGC\,1068 and NGC\,4579, respectively.
\item Circumnuclear rings are found in three inactive galaxies (NGC\,3351 Fig.~\ref{pair5}, NGC\,5248 Fig.~\ref{pair3} and NGC\,5806 Fig.~\ref{pair2}) and the Seyfert 2 NGC\,6951 (Fig.~\ref{pair7}), which correspond with known ring-like star forming structures in the central regions of these four galaxies \citep[][respectively]{3351_1, 5248_1, 5806_1}. 
\item In five other galaxies, asymmetric structures are observed. The emission line distribution of NGC\,4151 is  elongated from the centre to the South-West side of the FOV (Fig.~\ref{pair4}) in agreement with the high-excitation emission line feature described by \cite{4151_perez_89}, corresponding to the Extended Narrow Line Regions (ENLR) of this galaxy. The Seyfert 2 galaxy NGC\,2655 (Fig.~\ref{pair1}) presents a hot-spot 15$\arcsec$ away from the centre to the East and a lane 10$\arcsec$ West of the centre elongated along the South/North direction consistent with the polar ring observed by \citet{2655_2}. Asymmetric and irregular structures observed in the ionised gas intensity maps of NGC\,3627 (Fig.~\ref{pair2}) or NGC\,5055 (\ref{pair6}) are certainly due to the presence of dust in the circumnuclear regions. In the case of NGC\,5194 both the stellar continuum and the \Hb\ emission line intensity present irregularities consistent with the dusty nuclear spiral \citep{peeples_06}, while \OIII\ emission lines traces an outflow structure associated with the AGN. 
\item Finally three galaxies (two inactive NGC\,2985, Fig.~\ref{pair4}, NGC\,4459, Fig.~\ref{pair1} and the Seyfert 1 NGC\,4051, Fig.~\ref{pair3}) show regular round gaseous distributions.
\end{itemize}
%Morphologies of the \Hb\ and \OIII\ emission lines are generally very similar, except for NGC\,5194, its control NGC\,5055 (Fig.~\ref{pair6}) and NGC\,1068 (Fig.~\ref{pair0}). The \Hb\ emission line is distributed all over the FOV without specific features for the paired galaxies NGC\,5194 and NGC\,5055. The \OIII\ distribution shows an asymmetric structure elongated North of the nucleus for NGC\,5194, tracing the outflow structure \citep{M51_bubble}, and a V-shaped structure along the North/South direction is present in the \OIII\ emission line intensity map of NGC\,5055. For NGC\,1068 the distribution of the \OIII\ emission is very asymmetric. It is found predominantly North-East of the nucleus, tracing the ionisation cone, while \Hb\ intensity traces the nuclear spiral \citep{1068_1}.

\subsubsection{Ionised Gas Line Ratios}
As expected, the Seyfert galaxies present higher \OIII/\Hb\ ratios than their associated inactive galaxies in the central kpc regions. The maximum measured ratio for the inactive galaxies is of the order of 2, while for Seyfert galaxies it can reach values up to 10 or 20. For the Seyfert galaxies, the emission line ratio reaches its highest value in the central few arcseconds associated with the active nucleus. Some regions away from the central engine are also characterised by high \OIII/\Hb\ values ($\gg 1$) corresponding to high excitation regions, such as in the ionisation cone of NGC\,1068 (Fig.~\ref{pair0}), the bubble of NGC\,5194 (Fig.~\ref{pair6}) or the elongated gaseous feature of NGC\,4151 (Fig.~\ref{pair4}). Finally the ring-like structures seen in the gaseous maps of NGC\,3351, NGC\,5248 and NGC\,5806 show low emission lines ratios (\OIII/\Hb\ $\lesssim 0.1$) consistent with  star formation.

\subsection{Stellar and Ionised Gas Kinematics}
Outside the regions dominated by non-gravitational motions associated with AGN-driven outflows, all of our galaxies show stellar and gas velocity fields with a global regular rotation pattern (second panel of each row in Figs.~\ref{pair0} to \ref{pair7}). More complex dynamical structures are observed in some cases, especially in the ionised gas velocity fields. 
\label{sec:result_kin}
\subsubsection{Stellar Kinematics}

Most of the galaxies in our sample present a regular stellar velocity field, the orientation of their kinematic major-axis being constant over the FOV and their minor-axis perpendicular to it (e.g. NGC\,2985 Fig.~\ref{pair4}, NGC\,4459 Fig.~\ref{pair1}, NGC\,5055 Fig.~\ref{pair6}). A few active galaxies show some departures from axisymmetry e.g. the S-shaped zero-velocity line in NGC\,1068 (Fig.~\ref{pair0}) and twisted kinematic major-axis in NGC\,3627 (Fig.~\ref{pair2}). In all of our sample galaxies, the stellar kinematic major-axis is rather well aligned with the outer disc photometric major-axis (see Sect.~\ref{sec:PAphotPAkin}). The observed stellar kinematics within the \Sauron\ field of view are also clearly dominated, in all cases, by rotational motion, as evaluated via a global measurement of $V/\sigma$  (Dumas et al., in preparation), confirming the disc-like nature of the galaxies  in our sample.

For some of our galaxies, the stellar velocity dispersion decreases inwards. These so-called $\sigma$-drops \citep{eric_2001, Marquez2003} are observed in at least five Seyfert galaxies NGC\,1068, NGC\,2655, NGC\,3227, NGC\,4051, NGC\,4151 and NGC\,6951 and  one inactive galaxy NGC\,5248. These central $\sigma$-drops have been found to be common in S0 to Sb spiral galaxies \citep{paper3,paper7,Marquez2003} and recently \citet{ganda_2006} showed that such structures are present in galaxies as late as Sd. Central velocity dispersion drops are thought to be associated with dynamically cold structure, like discs and star formation regions \citep{wozniak_2003}.

\subsubsection{Ionised Gas Kinematics}

The ionised gas velocity maps are dominated by rotation and for all of the galaxies the gas rotates faster than the stars (Figs.~\ref{pair0} to \ref{pair7} second panels of the second rows). Since stars follow collisionless orbits, their mean velocities are indeed expected to be lower than the local circular velocity \citep[see][]{BinneyTremaine}.  The gas velocity fields are more distorted and present richer structures than the stellar ones. Evidence for significant deviations from axial symmetry exist in almost all the galaxies: radial change of orientation of the major-axis (NGC\,3227, Fig.~\ref{pair0}), wiggles along the kinematic minor-axis (NGC\,2655, Fig.~\ref{pair1} , NGC\,4579, Fig.~\ref{pair5}), S-shaped feature (NGC\,1068, Fig.~\ref{pair0}), or more complicated structures (e.g. NGC\,3627, Fig.~\ref{pair2} or NGC\,5194, Fig.~\ref{pair6}  ).  For all of our galaxies, the \Hb\ and \OIII\ velocity fields present very similar structure, though \OIII\ velocity fields have generally better signal-to-noise ratio than the \Hb\ ones, except for NGC\,4459 (Figs.~\ref{pair0} and \ref{kinHb1}). We also observe highly misaligned kinematic major-axes of the ionised gas with respect to the stellar ones in some of our galaxies such as NGC\,2655 (Fig.~\ref{pair1}) and NGC\,4579 (Fig.~\ref{pair5}). These misalignments will be analysed in the following Section.

The gaseous velocity dispersion maps present no particular structure for all of the inactive galaxies and two Seyferts (NGC\,3627, Fig.~\ref{pair2} and NGC\,4051, Fig.~\ref{pair3}). For eight Seyfert galaxies (NGC\,1068, NGC\,2655, NGC\,3227, NGC\,3627, NGC\,4151, NGC\,4579, NGC\,5194 and NGC\,6951) and three inactive galaxies (NGC\,2985, NGC\,4459 and NGC\,5055) $\sigma$ values increase inwards. In the inner 5$\arcsec$, the dispersion can reach values 200 km~s$^{-1}$ higher than in the outer part of the FOV as in NGC\,1068 (Fig.~\ref{pair0}). NGC\,2655 (Fig.~\ref{pair1}) and NGC\,4579 (Fig.~\ref{pair5}) also present high dispersion values ($\sim$ 230 km~s$^{-1}$) associated with the gas distribution structures. For NGC\,4151, gaseous dispersion velocity maps exhibit high values in the inner arcseconds (about 280 km~s$^{-1}$) corresponding to the central engine. Outside the very inner part and inside the ENLR, the velocity dispersion shows lower values (between 50 and 130 km~s$^{-1}$). Then, $\sigma$ increases in the outside parts of the map for this galaxy, corresponding to the location of a dusty ring \citep{4151_dust}.  

%In the following, the gaseous kinematic analysis was done on the \OIII\ component, except for NGC\,4459, for which \Hb\ was used .

\begin{subfigures}
\begin{figure*}
\centering
\includegraphics[height=20cm ]{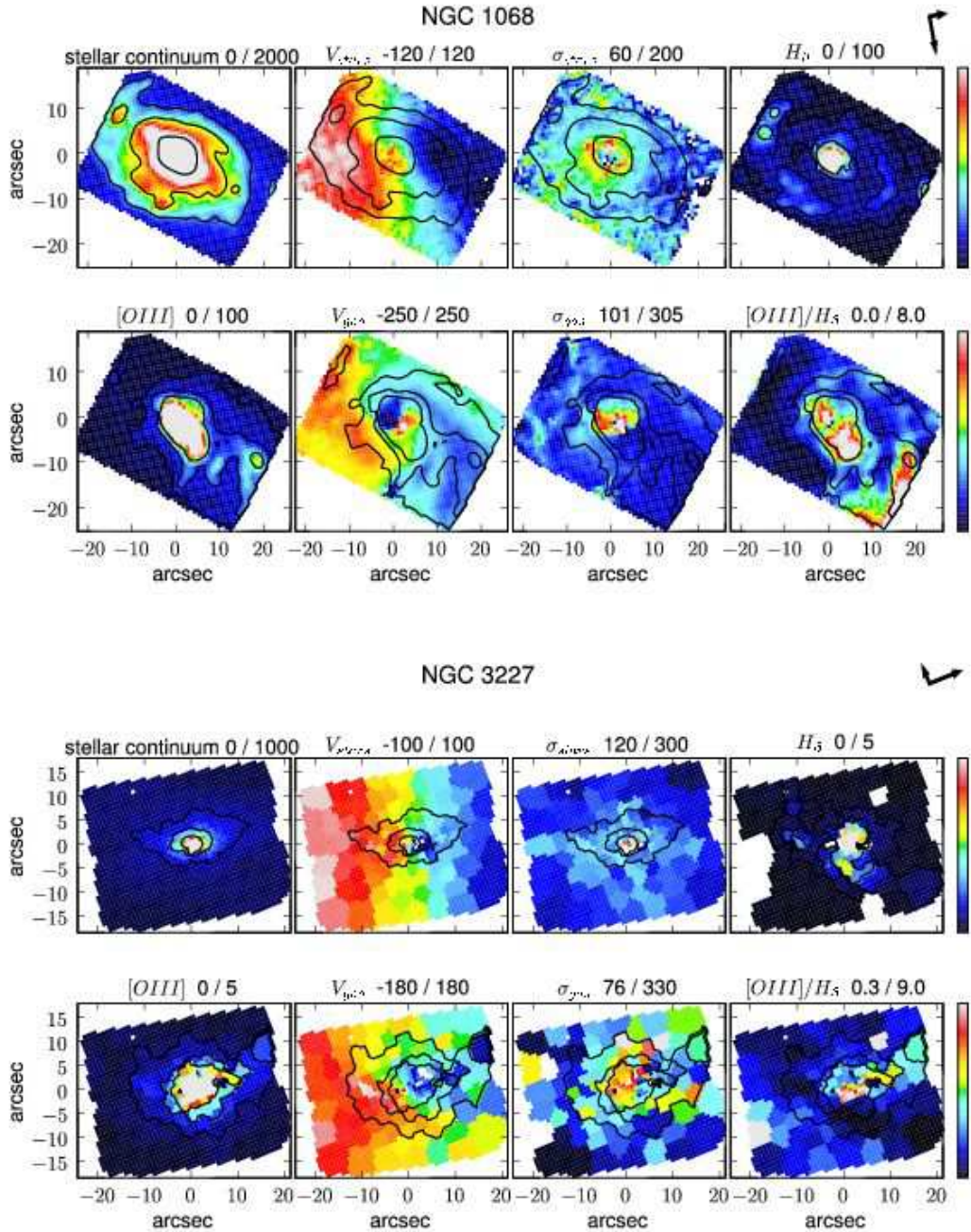}
\caption{\Sauron\  maps for the Seyfert galaxies NGC\,1068 (top panels) and NGC\,3227 (bottom panels). For each galaxy the first row shows (from left to right) the reconstructed continuum map, the stellar velocity field, the stellar velocity dispersion map and \Hb\ intensity distribution. The second row consists of the \OIII\ intensity velocity and velocity dispersion maps and the line ratio \OIII/\Hb\ map. All the intensity maps are in units of 10$^{-16}$ erg s$^{-1}$ cm$^{-2}$. The stellar and emission lines intensity maps are overlaid on the stellar and ionised gas kinematics.  Velocities and velocity dispersions are in km\,s$^{-1}$. The color scale is shown on the right hand side and the cut levels are indicated in the top right corner of each map. All the maps are orientated so that the photometric major-axis of the galaxy is on the x-axis. The long and short arrows on the right of each galaxy name show the North and the East directions, respectively.}% and the scale is shown in the first panel.}.
\label{pair0}
\end{figure*}
\begin{figure*}
\centering
\includegraphics[height=22cm ]{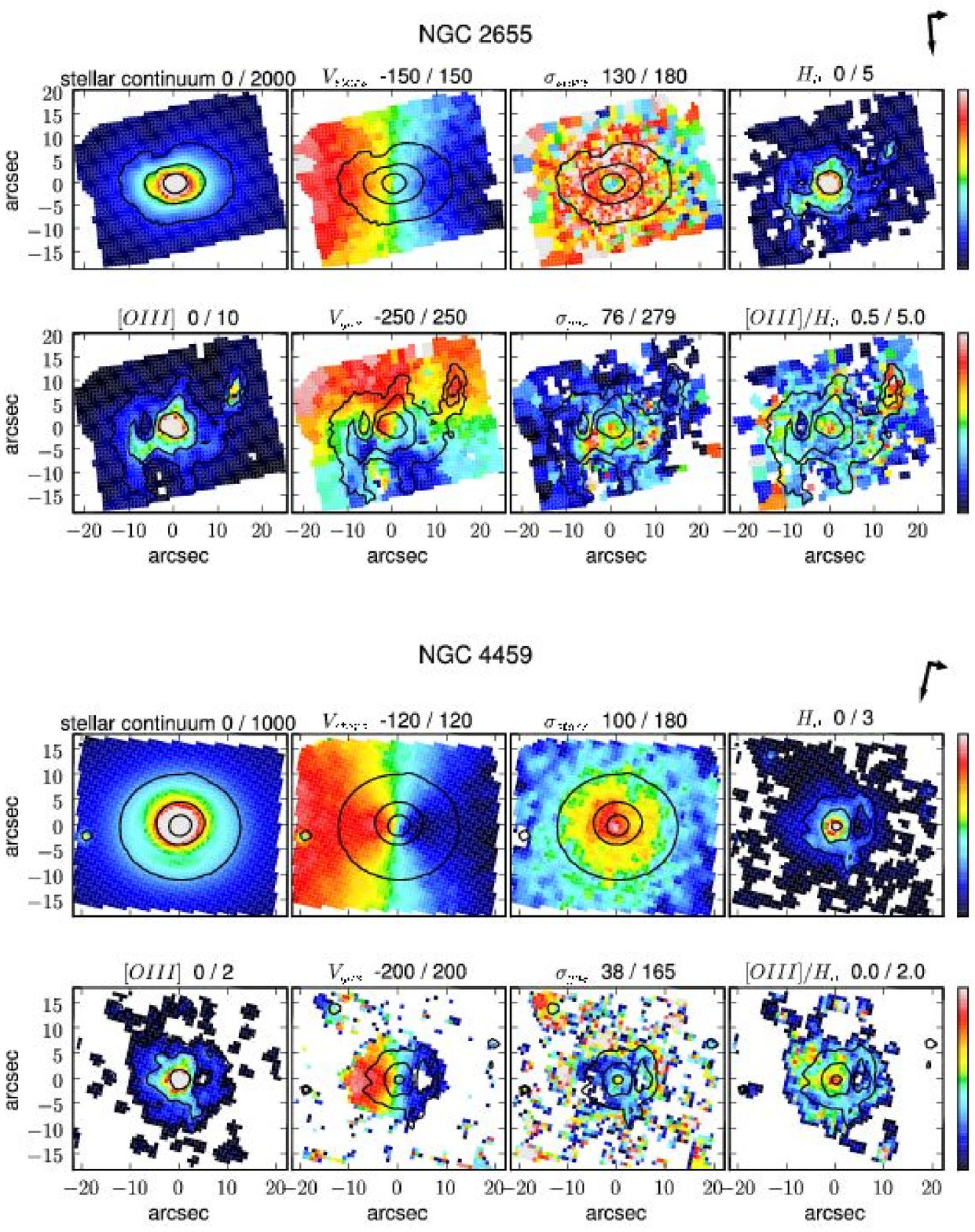}
\caption{\Sauron\  maps for NGC\,2655 (top panels)  and its control NGC\,4459 (bottom panels). See caption of Fig.~\ref{pair0} for details.}
\label{pair1}
\end{figure*}
\begin{figure*}
\centering
\includegraphics[height=22cm ]{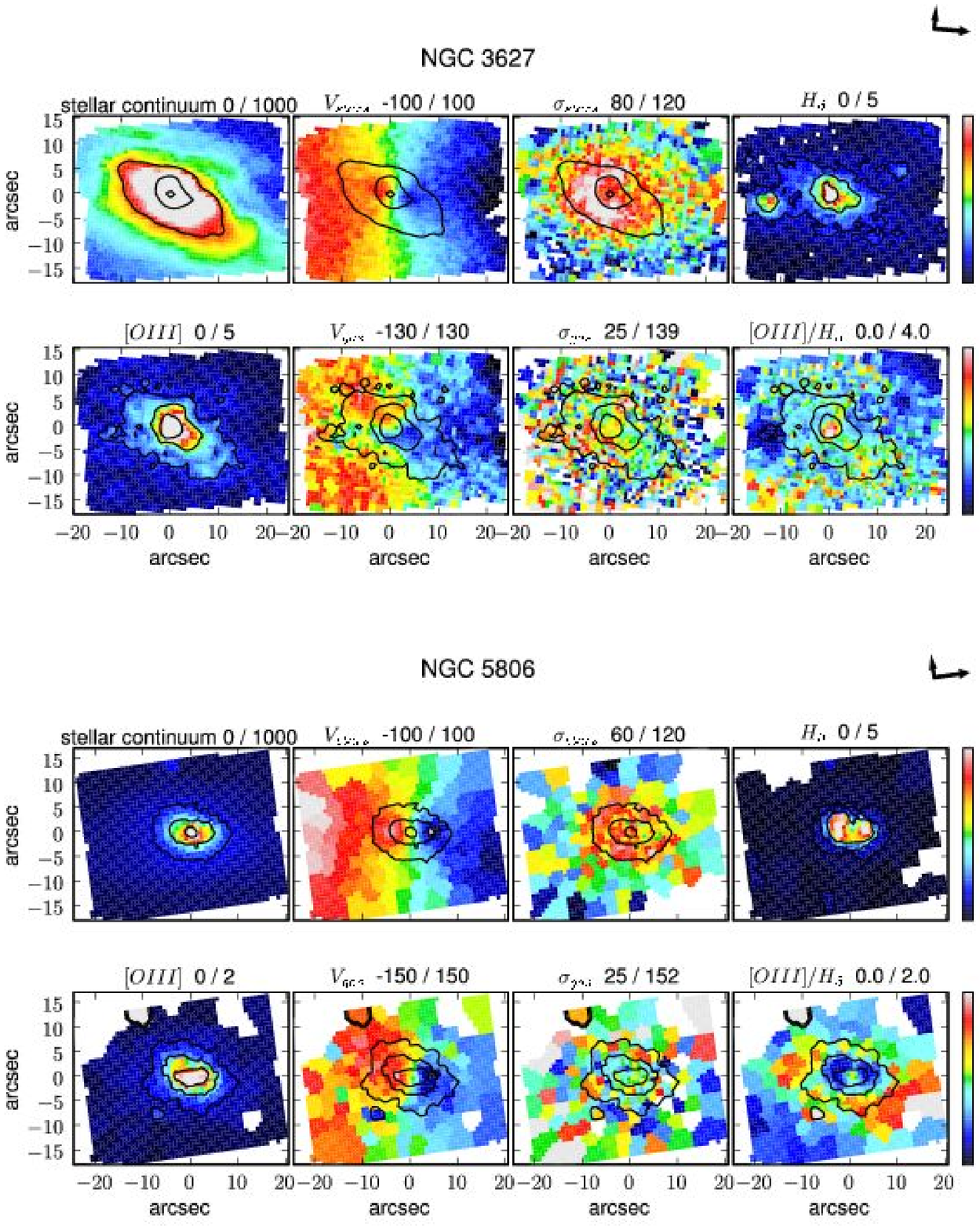}
\caption{\Sauron\  maps for NGC\,3627 (top panels)  and its control NGC\,5806 (bottom panels). See caption of Fig.~\ref{pair0} for details. }
\label{pair2}
\end{figure*}
\begin{figure*}
\centering
\includegraphics[height=22cm ]{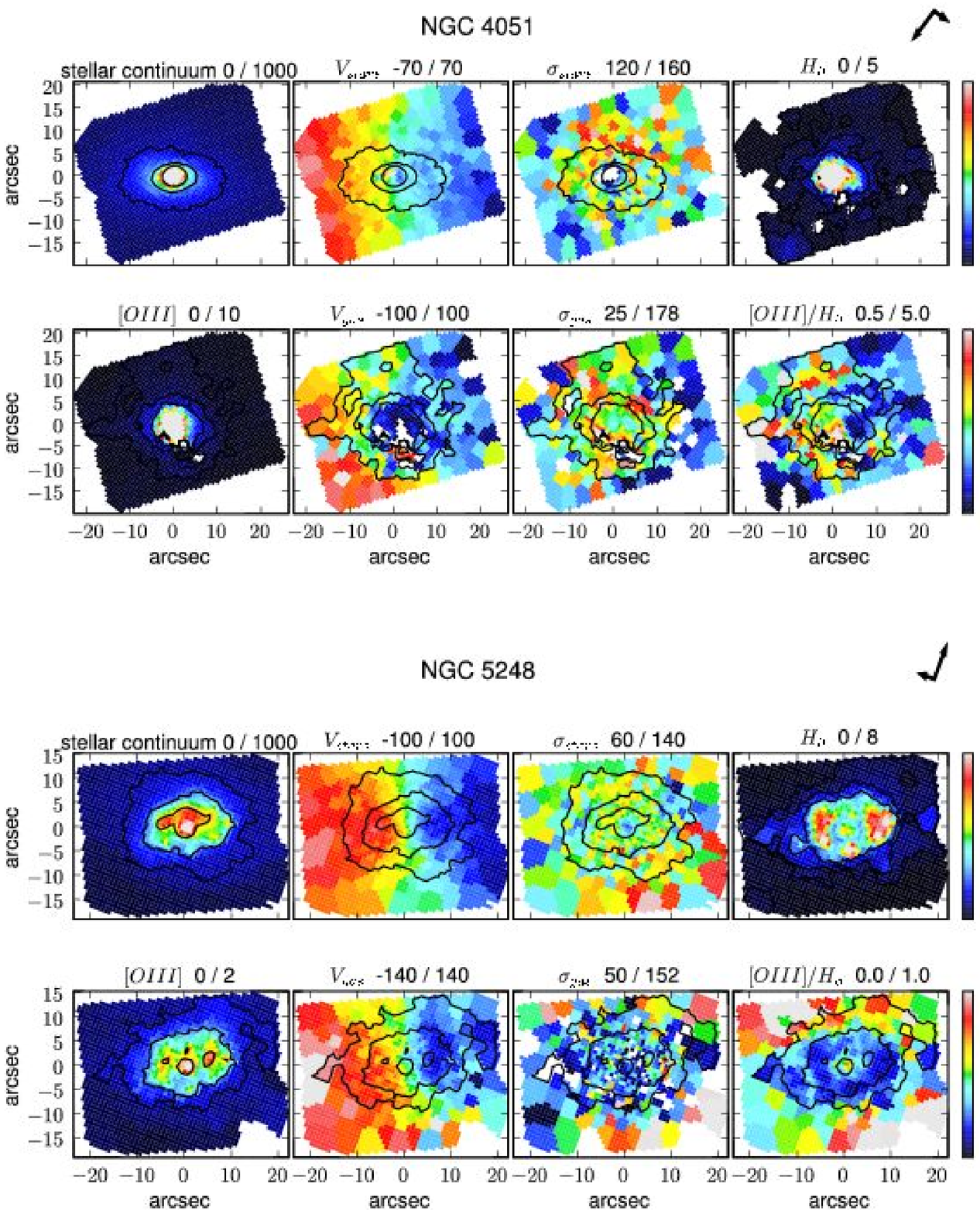}
\caption{\Sauron\  maps for NGC\,4051 (top panels)  and its control NGC\,5248 (bottom panels). See caption of Fig.~\ref{pair0} for details.}
\label{pair3}
\end{figure*}
\begin{figure*}
\centering
\includegraphics[height=22cm ]{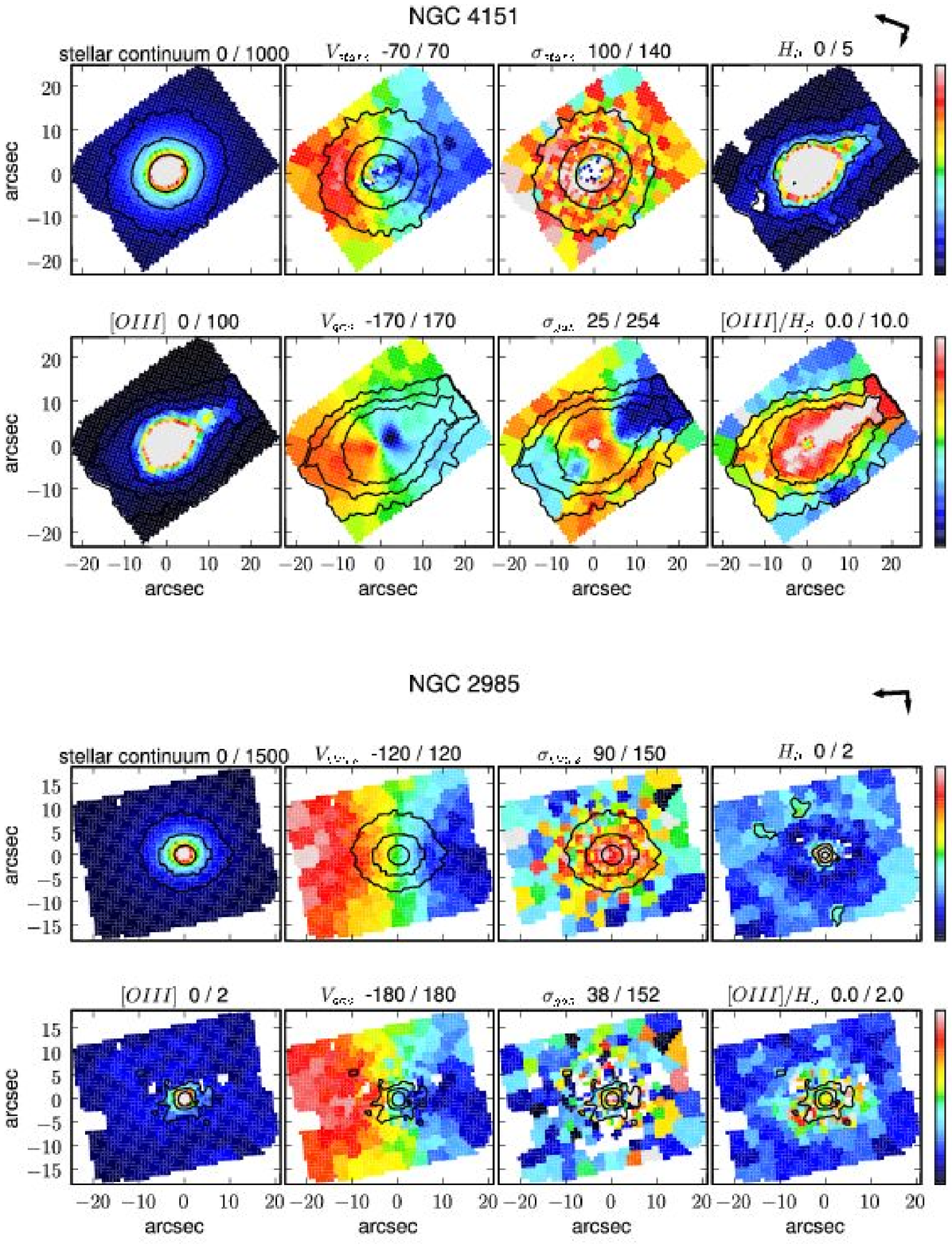}
\caption{\Sauron\ maps for NGC\,4151 (top panels)  and its control NGC\,2985 (bottom panels). See caption of Fig.~\ref{pair0} for details. }
\label{pair4}
\end{figure*}
\begin{figure*}
\centering
\includegraphics[height=22cm ]{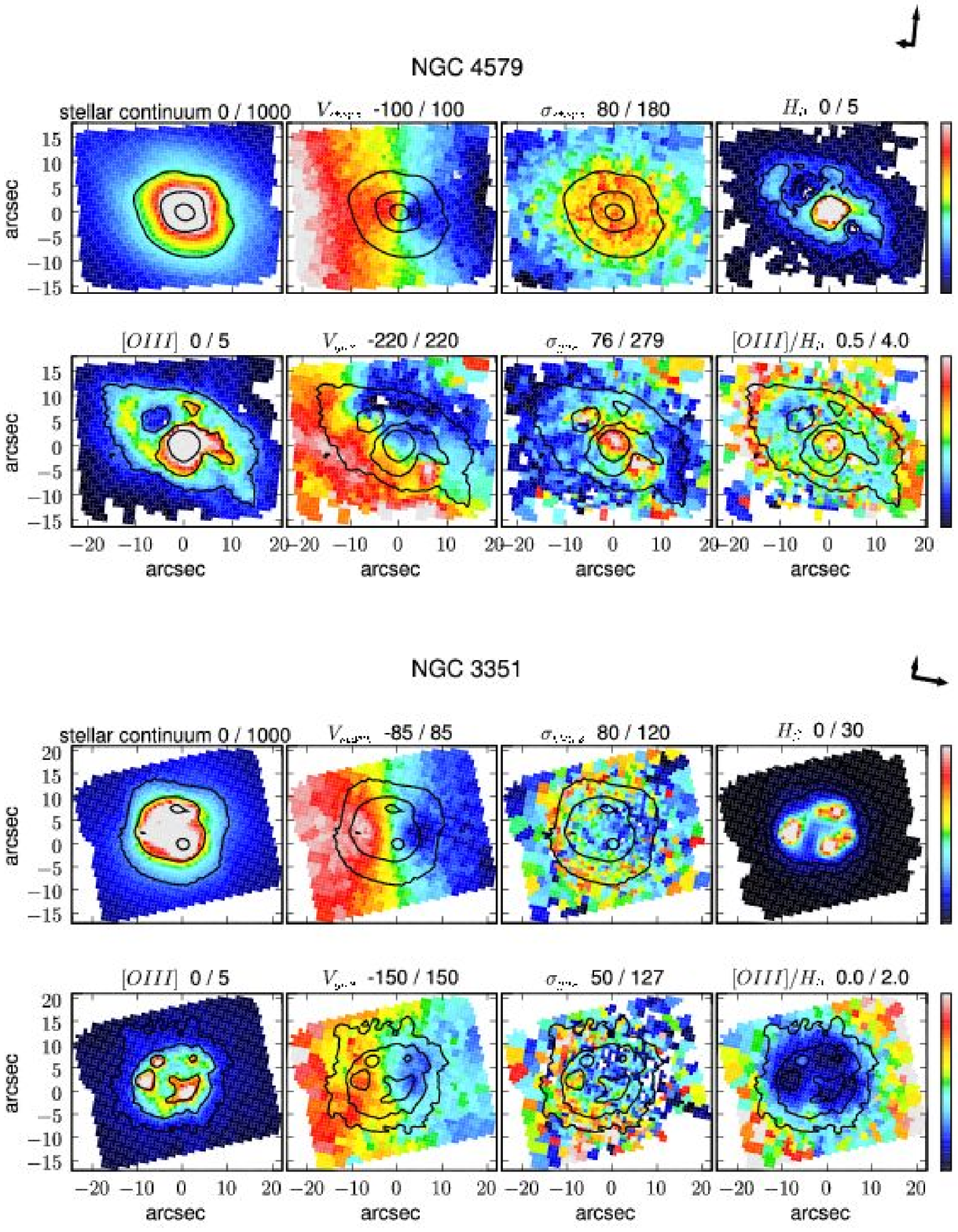}
\caption{\Sauron\ maps for NGC\,4579 (top panels)  and its control NGC\,3351 (bottom panels). See caption of Fig.~\ref{pair0} for details.} 
\label{pair5}
\end{figure*}
\begin{figure*}
\centering
\includegraphics[height=22cm ]{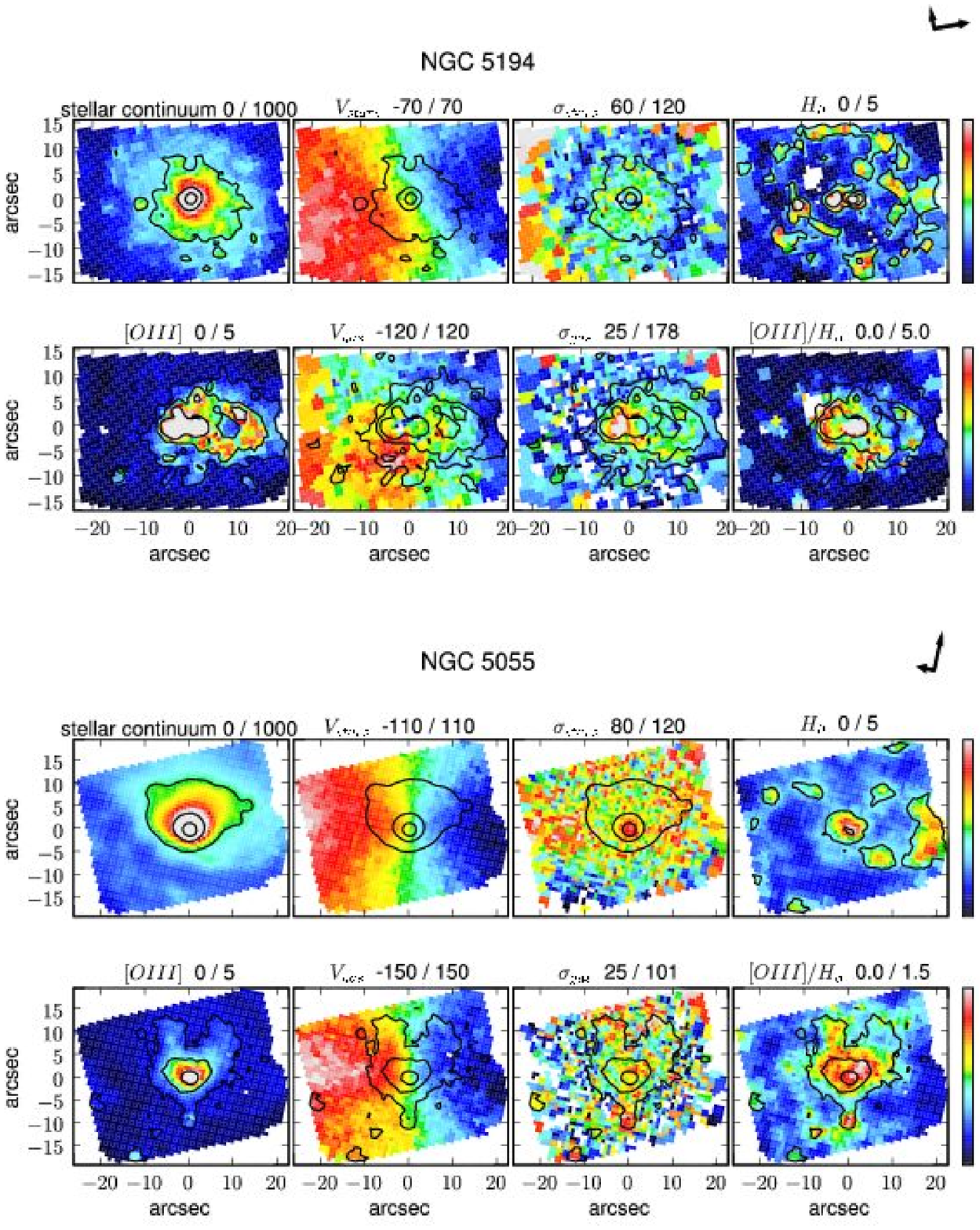}
\caption{\Sauron\ maps for NGC\,5194 (top panels)  and its control NGC\,5055 (bottom panels). See caption of Fig.~\ref{pair0} for details. }
\label{pair6}
\end{figure*}
\begin{figure*}
\centering
\includegraphics[height=22cm ]{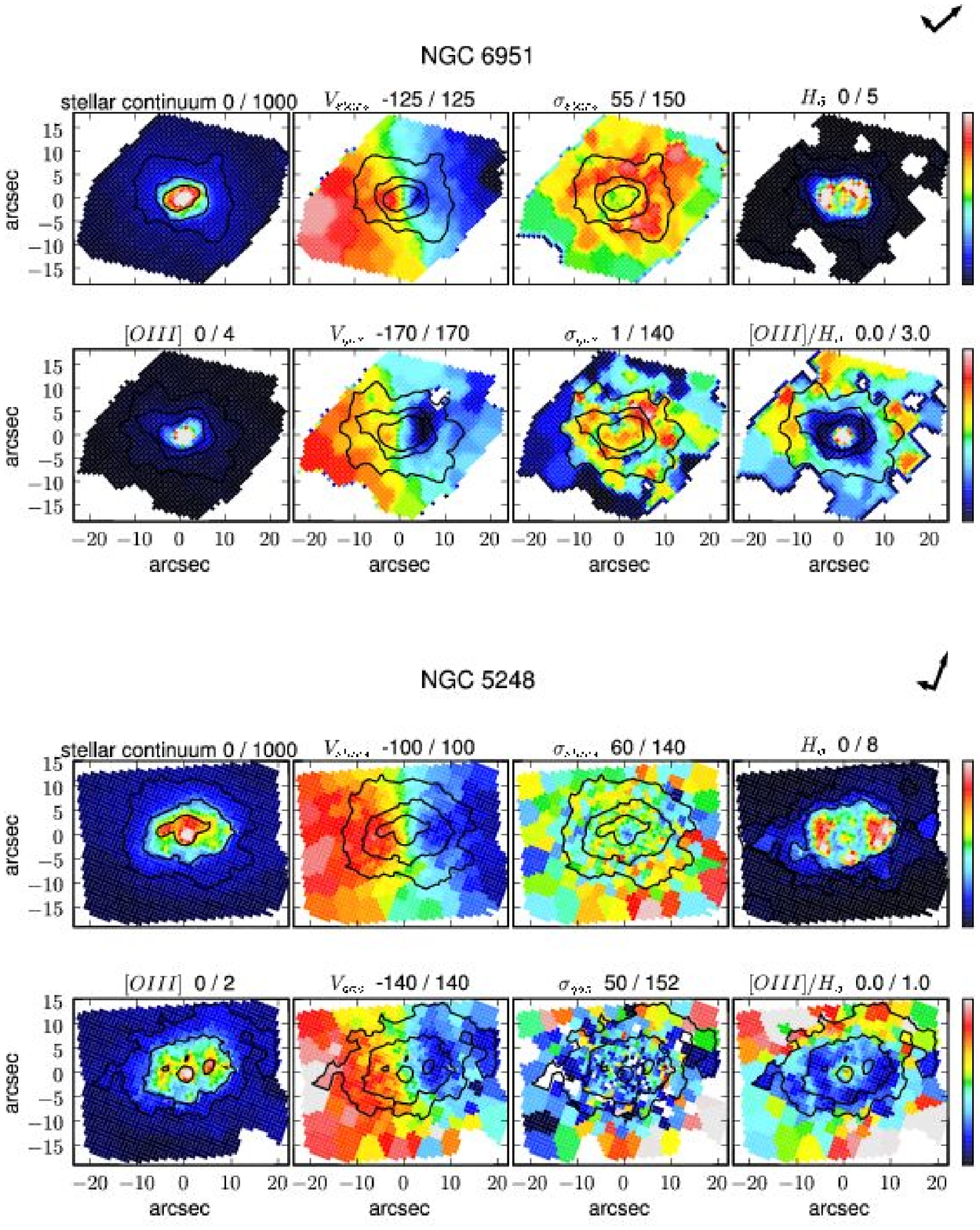}
\caption{\Sauron\ maps for NGC\,6951 (top panels)  and its control NGC\,5248 (bottom panels). See caption of Fig.~\ref{pair0} for details. }
\label{pair7}
\end{figure*}
\end{subfigures}

%% file: global_kinematic.tex
\section{Comparison of Stellar and Gaseous Kinematics}
\label{sec:ana}

In the previous Section we described the \Sauron\  maps of the stellar and ionised gas components for the Seyfert and inactive galaxies. For all of our galaxies, the stellar velocity fields are dominated by rotation, showing regular isovelocity contours and their photometric and kinematic major-axes seem globally aligned. In comparison with the stellar maps, however, the gaseous distributions are more complex and the kinematics more perturbed with this effect appearing more pronounced for the Seyferts compared to their control galaxies. The identification of this qualitative difference between gaseous and stellar distributions and kinematics in the inner kpc of active and inactive galaxies has important implications for the triggering and nuclear fueling.
In this Section, we therefore provide a more quantitative analysis of the stellar and gaseous velocity fields and the photometric major-axis of the galaxies. We first compare the major-axis orientation of the stellar velocity field to that of the outer galactic disc. Then we quantify the differences between stars and gas in Seyfert and inactive galaxies as seen in our \Sauron\  maps. We compare the global orientation of the stellar and ionised gas kinematic maps to probe any significant differences between the two components. Then we analyse the kinematic maps in more detail, by computing the kinematic parameters (systemic velocity, dynamical centre, PA, rotational velocity) as a function of radius, using  a simple representation of a two-dimensional thin disc in rotation.

\subsection{Global Kinematic Major-Axis Orientation}
\label{sec:gpa}
\begin{figure}
\centering
\includegraphics[width=8.4cm]{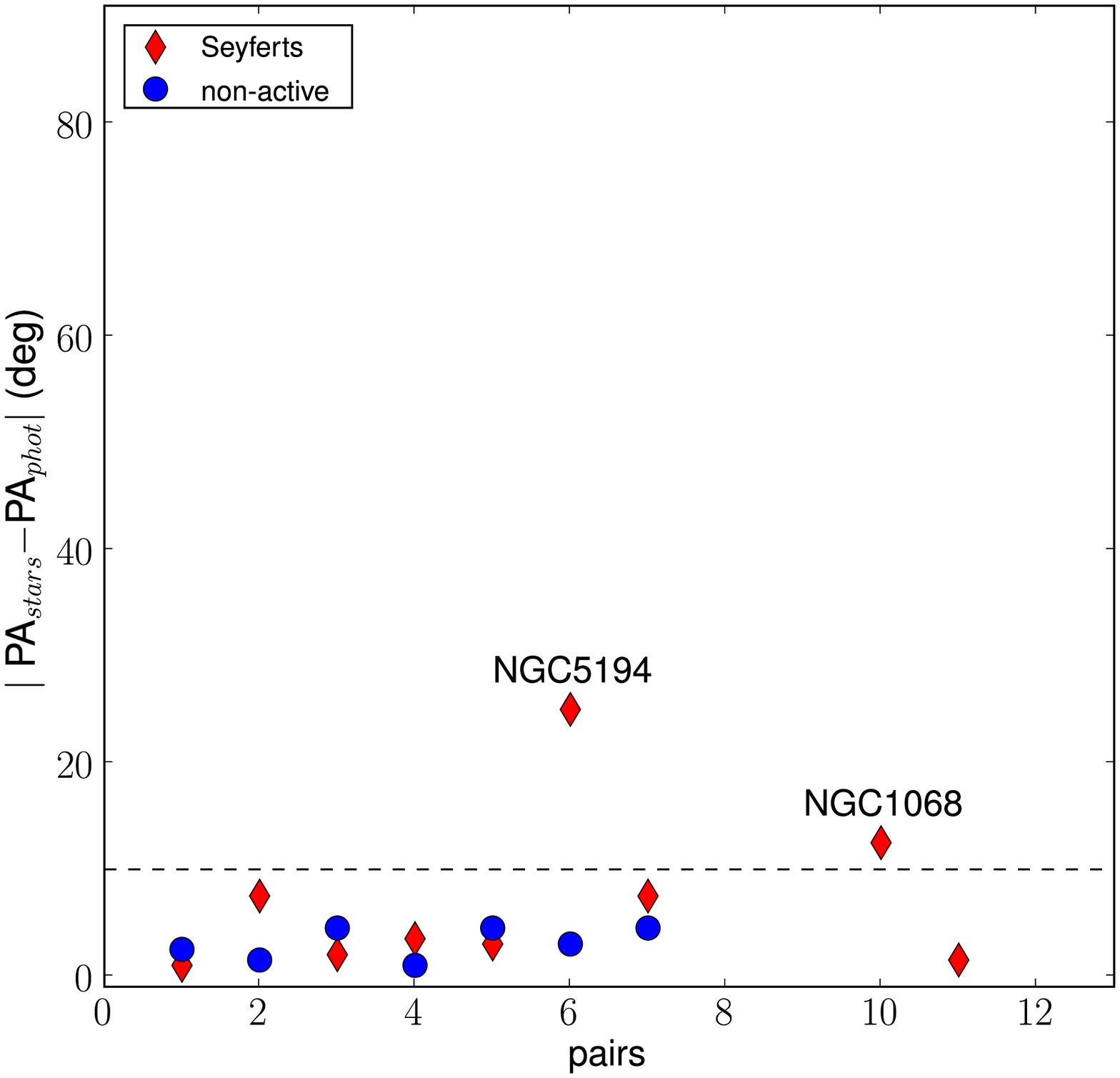}
\caption{Distribution of the differences between the photometric and the global stellar kinematics PAs  $PA_{stars}-PA_{phot}$ for each pair of galaxies, Seyferts (red diamonds) and inactive galaxy (blue circles). The dashed black horizontal line corresponds to a difference of $10\degr$.}
\label{PAs_PAp}
\end{figure}

The stellar and gaseous velocity fields can be first simply parameterised by their major-axes.  Misalignments between the kinematic major-axis of the stellar and ionised gas components may reveal the presence of non-axisymmetric structures or decoupled components such as central discs.  Here, we compare such misalignments in our sample of Seyfert and control galaxies,  to characterise global kinematic differences in their central kpc.

%The angular momenta of gas and stars tend to become parallel or anti-parallel in the first case, and orthogonal in the second one. Therefore in the case of an external origin for the gas we except a distribution of kinematic misalignments between the stars and gas around 0, 90 and 180$\degr$. It has been demonstrated for early type galaxies \citep{paper5} that the values of the kinematic misalignment between the stellar and gaseous components do not depend on Hubble type, galactic environment and galaxy luminosity. Here we compare such misalignments in Seyfert and control galaxies, in order to search for kinematic differences in the central kpc regions between these 2 distributions.\\

The kinematic axis orientations were determined for both the stellar and ionised gas velocity fields. The global position angles (PAs) of the kinematic major-axis were obtained by minimizing the differences between our \Sauron\  velocity fields (stars and \OIII) and a bi-antisymmetric representation of these fields. This method is detailed by \citet{kinemetry} in their Appendix C and we used a specific implementation written by Michele Cappellari. The values for the measured kinematics PAs and misalignments are listed in Table \ref{table:pas}.

\subsubsection{Stellar Kinematic and the Orientations of the Line of Nodes}
\label{sec:PAphotPAkin}

In the case of an axisymmetric mass distribution, the direction of the stellar kinematic major-axis should coincide with the line-of-nodes (LON) of the galaxy, while in the case of a triaxial potential the kinematic and photometric major-axes can depart from each other due to projection effects. We first therefore check if the stellar kinematic major-axis in the central regions and the outer photometric major-axis are aligned. The photometric major-axis position angle is computed from ellipse fitting on the R-band DSS images for each of our galaxies. These derived values are then compared with published ones. For most of our galaxies, the PAs values found in the literature are in good agreement with those derived from the ellipse fit, within the error bars, in which case we took our measured values as the photometric PAs. However, a few galaxies present significant discrepancies between our fitted values and the published ones. This is the case for NGC\,1068 and NGC\,4151 which present a weak oval bar and a faint, almost circular outer disc, respectively. The measurement of the photometric PA of NGC\,5194 is uncertain given that this galaxy is nearly face-on (i=20$\degr$). We therefore used reliable values derived from H{\sc i} kinematics for NGC\,4151 \citep{4151_Pedlar_92} and optical kinematics for NGC\,5194 \citep{tully_M51},  as the photometric PAs. Table \ref{table:pas} lists the resultant photometric PA values for the sample galaxies. The differences between the global stellar kinematic and the photometric PAs are plotted in Fig. \ref{PAs_PAp}. 

Our first qualitative impression is confirmed: the stellar kinematic PAs are parallel to the photometric ones within 10$\degr$ for all of our galaxies except for NGC\,5194 and NGC\,1068 which present a difference of about 20$\degr$ and 15$\degr$, respectively. In the case of NGC\,1068, the stellar kinematic major-axis PA corresponds to the average PA of the kinematic axis as fitted on the large-scale H{\sc i} data by \citet{brinks_1997}, while the outer photometric major-axis lies at a PA of about 80\degr \citep{1068_1}.  Besides the two specific cases of NGC\,5194 and NGC\,1068, for all of our galaxies the major-axis orientation of the central part of the stellar velocity field is therefore a reliable measurement for the line-of-nodes of the galaxy.

\subsubsection{Global Kinematic Misalignments between Stars and Gas}

\begin{figure*}
\centering
\includegraphics[width=\textwidth]{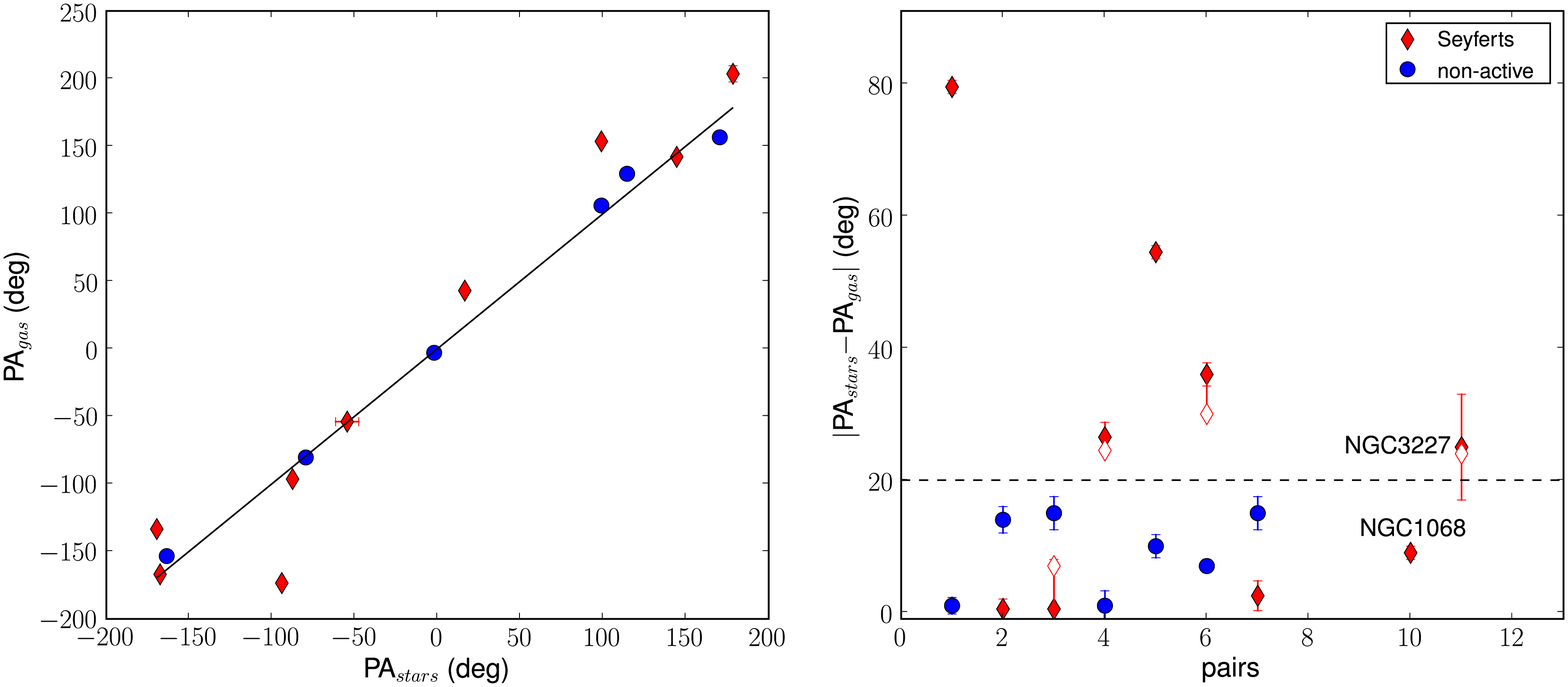}
\caption{Kinematic misalignments between stars and gas. Left panel: global gaseous versus stellar kinematic PAs. The one-to-one relation is shown as a black line, active galaxies and inactive galaxies are shown as filled red diamonds and filled blue circles respectively. Right panel: distribution of differences of global kinematics PAs of stars and gas $PA_{stars}-PA_{gas}$ for each pair of galaxies, Seyfert (red diamonds) and inactive galaxies (blue circles). The non-filled red diamonds correspond to the PA derived by masking the BLR of the Seyfert 1 galaxies NGC\,3227, NGC\,4051 (pair 3) and NGC\,4151 (pair 4) and the outflow bubble region for NGC\,5194 (pair 6). The dashed black horizontal line corresponds to a kinematic misalignment of $20\degr$.}
\label{global_pa}
\end{figure*}

Fig. \ref{global_pa} presents the global gaseous versus stellar kinematic PAs and the distribution of kinematic misalignments between the stellar and gaseous components in our sample. Most galaxies in our sample follow the simple one-to-one relation (left panel). Five Seyfert galaxies  show an absolute misalignment larger than 20$\degr$, and for all the inactive galaxies, the misalignment between gaseous and stellar kinematic major-axes is less than $20\degr$. Among the 5 Seyfert galaxies with strong kinematic misalignments, two are Seyfert 1 galaxies (NGC\,3227 and NGC\,4051) for which the BLR is detected and one is NGC\,5194 which presents an outflow structure in the gaseous maps  on the Northern side (Fig \ref{pair6}). The other two galaxies are NGC\,2655 and NGC\,4579. NGC\,2655 is known to host off-planar gas in the central regions \citep{2655_1} and an inner molecular spiral is present in the nuclear region of NGC\,4579 \citep{nugaIV}, a spiral-like structure is also observed in our ionised gas distribution maps (Fig. \ref{pair5}). 
In order to take out any influence of the AGN itself on the result of our analysis, we masked the BLR for the Seyfert 1 galaxies and the prominent outflow region of NGC\,5194 in the stellar and gaseous velocity fields. Then we determined new values for the kinematic position angles of the stars and gas. These values of the kinematic misalignment are listed in Table \ref{table:pas} for these four galaxies, and we show them as non filled red diamonds in the right panel of Fig. \ref{global_pa}. These new values are different from the PAs determined without masking the regions where the dynamics is dominated by the AGN. For NGC\,4151 and NGC\,5194, the kinematic misalignment between the stellar and ionised components is smaller after masking such regions. However, it is the opposite effect for NGC\,4051: the difference between the stellar and gaseous kinematic major-axis orientation is larger after masking the BLR. For NGC\,3227, there is no significant difference between the two values. After this second analysis, the same five Seyfert galaxies have still kinematic misalignments larger than 20$\degr$. 

Although about half of our Seyferts sample presents strong misalignments between stellar and gaseous kinematics while none of the inactive galaxies shows such differences, no statistically-significant difference is found between Seyfert and control-galaxies. A Mann-Whitney U test shows that the two distributions are identical. By symmetrizing the velocity fields to look for global differences in disc orientation, this zeroth order analysis shows that our sample size is too small to be able to separate statistically the two distributions of Seyfert and inactive galaxies, despite the fact that half of the Seyferts present kinematic misalignments larger than 20$\degr$.

\begin{table*}
\begin{center}
\begin{tabular}{|l|l|l|c|c|c|c|c|c|}
\hline
Pairs& Name & Activity 		&PA$_{phot}$	&Ref	&PA$_{stars}$	&PA$_{gas}$	&$\Delta$PA$_{phot-kin}$&$\Delta$PA$_{kin}$ \\
	&NGC	&		&(deg)		&	&(deg)		&(deg)		&(deg)		&(deg)\\
(1)	&(2)	&(3)		&(4)		&(5)	&(6)		&(7)		&(8)		&(9)\\
\hline
	&1068	&S2		&-100$\pm$5	&1	&-88$\pm$1	&-97$\pm$1	&12$\pm$6	&9$\pm$2\\
\cline{2-9}
	&3227	 &S1		&158$\pm$2	&2	&160$\pm$5	&184$\pm$1	&2$\pm$7	&24$\pm$6\\
\hline
1	&2655	 &S2		&-95$\pm$5	&3	&-94$\pm$1	&-174$\pm$1	&1$\pm$6	&80$\pm$2\\
	&4459	 &inactive	&-77$\pm$2	&3	&-80$\pm$1	&-81$\pm$1	&3$\pm$3	&1$\pm$2\\
\hline
2	&3627	 &S2		&-175$\pm$5	&3	&-168$\pm$1	&-167$\pm$1	&7$\pm$6	&1$\pm$2\\
	&5806	 &inactive	&172$\pm$2	&3	&171$\pm$2	&157$\pm$1	&1$\pm$4	&14$\pm$3\\
\hline
3	&4051	 &S1		&-45		&4	&-47$\pm$3	&-54$\pm$1	&6$\pm$3	&7$\pm$4\\
	&5248	 &inactive	&110$\pm$2	&3	&115$\pm$2	&130$\pm$1	&5$\pm$4	&15$\pm$3\\
\hline
4	&4151	 &S1		&22		&5	&19$\pm$3	&43$\pm$1	&1$\pm$3	&24$\pm$4\\
	&2985	 &inactive	&-3$\pm$1	&3	&-2$\pm$2	&-3$\pm$1	&1$\pm$3	&1$\pm$3\\
\hline
5	&4579	 &S2		&96$\pm$2	&3	&99$\pm$1	&154$\pm$1	&3$\pm$5	&55$\pm$2\\
	&3351	 &inactive	&-168$\pm$2	&3	&-164$\pm$1	&-154$\pm$1	&6$\pm$3	&10$\pm$2\\
\hline
6	&5194	 &S2		&-190		&6	&-165$\pm$5	&-134$\pm$1	&25$\pm$2	&30$\pm$6\\
	&5055	 &inactive	&102$\pm$2	&3	&99$\pm$1	&106$\pm$1	&3$\pm$4	&7$\pm$2\\
\hline
7	&6951	 &S2		&137		&7	&145$\pm$2	&142$\pm$1	&7$\pm$3	&3$\pm$3\\
	&5248	 &inactive	&110$\pm$2	&3	&115$\pm$2	&130$\pm$1	&5$\pm$4	&15$\pm$3\\
\hline
\end{tabular}
\end{center}

\caption{Results of the kinematic analysis of the stellar and gaseous components. (1): Pair identifier; (2): NGC\, number; (3): Activity; (4): Photometric position angle (PA) in degrees. The values for NGC\,4151 and NGC\,5194 correspond to the line of nodes PA. (5): Origin of the values for the photometric PA: 1: \citet{1068_1}, 2: \citet{3227_carole_95b}, 3: R-Band DSS images ellipse fitting, 4: \citet{4051_outflow}, 5: \citet{4151_Pedlar_92}, 6: \citet{tully_M51}, 7: \citet{6951_1}; (6): Stellar Kinematic position angle (PA), in degrees; (7): Gas Kinematic position angle (PA), in degrees; (8): Difference between the photometric and the stellar kinematic PAs, in degrees $\Delta$PA$_{phot-kin}=|$PA$_{phot}-$PA$_{stars}|$; (9): Difference between the stars and gas kinematic PAs, in degrees $\Delta$PA$_{kin}=|$PA$_{stars}-$PA$_{gas}|$. \label{table:pas}}
\end{table*}

\subsection{Radial Variations of Kinematic Properties}
\label{sec:radius}

 In the above analysis, global parameters were extracted from
the two-dimensional data, and therefore details of the observed
kinematic structures and small-scales variations of the PA are not
accounted for. Since we have the full two-dimensional dynamical information, we can analyse the stellar and ionised gas velocity fields quantitatively over the  \Sauron\  FOV and then study the kinematic properties as a function of radius. Fourier expansion of the velocity field along ellipses is a powerful technique to unveil the kinematic perturbations and therefore to give information on the gravitational potential \citep{schoenmakers_1997, Wong_2004, kinemetry}. This formalism will be used in a second paper (Dumas et al. in preparation) on our \Sauron\ data associated with large scale radio data obtained with the Very Large Array (VLA). For the current paper, we estimate the kinematic parameters of our stellar and gaseous velocity fields by modeling the galactic disc as a two-dimensional thin disc, in pure circular rotation with no vertical velocities. The line-of-sight velocities can then be written as:

%The above analysis extracted one dimensional information (the position angle) from two dimensional data, and thus kinematic structures and variations of the PA on small scales are not visible. 
\begin{eqnarray}
\label{vlos}
V_{los}(R,\Phi,i)=V_{sys} & + & V_{rot}(R)\cos\Phi \sin i \nonumber \\
                          & + & V_{rad}(R)\sin\Phi \sin i
\end{eqnarray}

where $V_{sys}$\ is the systemic velocity of the galaxy, $V_{rot}$\ and $V_{rad}$\ are the rotational and radial velocities in the galactic plane. They depend only on the radius R. The disc inclination $i$\ ranges from $0\degr$\ (face on) to $90\degr$\ (edge on). Finally, the polar coordinates in the plane of the galaxy $(R,\Phi)$\ are related to the observable coordinates $(x,y)$\ on the sky plane by:
\begin{equation}
\left\{ 
\begin{array}{l}
\cos\Phi = \frac{\displaystyle
	-(x-X_{c})\sin\Phi_0+(y-Y_{c})\cos\Phi_0}{\displaystyle R} \\
	\\
\sin\Phi = \frac{\displaystyle -(x-X_{c})\cos\Phi_0-(y-Y_{c})\sin\Phi_0}{\displaystyle R \cos i}
\end{array}
\right.
\end{equation}

$(X_c,Y_c)$\ are the coordinates of the dynamic centre and $\Phi_0$\ is the PA of the projected major-axis of the disc measured with respect to North (counter-clockwise).

We use the expression of the velocity (Eq.~\ref{vlos}) with the tilted-ring method in order to derive the stellar and gaseous kinematic parameters. Our technique is inspired from the ROTCUR routine in the GIPSY package \citep{rotcur}. The galactic disc is divided into concentric ellipses determined by 6 parameters~: the centre  $(X_c,Y_c)$, position angle $\Phi_0$, inclination $i$, the offset velocity $V_{sys}$ and the rotational velocity $V_{rot}$. For this first iteration, $V_{rad}$ is set to 0. The width of each ring is chosen to be large enough to have a sufficiently high number of pixels so the fit is good, and small enough to assume the rotational velocity is constant along the ring. The width of the rings is finally chosen to be equal to the seeing. All these parameters (except $V_{rad}$) are adjusted iteratively for each ring, by fitting the expression of the line-of-sight velocity of Eq. \ref{vlos} to the observed velocity field:

\begin{enumerate}
\item Initial values for $V_{sys}$, $\Phi_0$\ and $i$\ were taken from previous published studies or the NASA Extragalactic Database (NED). We fixed the dynamical centre to coincide with the photometric centre, or with the location of the central radio continuum peak for the Seyfert galaxies. 
\item A first fit is done, by fixing $\Phi_0$ and $i$ while allowing $V_{sys}$, $(X_c,Y_c)$ and $V_{rot}$ to vary as a function of R.  The mean values of $V_{sys}$, $X_c$ and $Y_c$ were taken as  improved estimates for the systemic velocity and the dynamical center of the galaxy. 
\item Then, we fit the position angle $\Phi_0$ and $V_{rot}$ with $V_{sys}$ and $(X_c,Y_c)$ fixed at the values derived from the previous step, and the disc inclination fixed at its initial guess
\end{enumerate}

These three steps were followed for the stellar velocity fields of our sample. For the gaseous velocity fields, we imported the values of $V_{sys}$ and $(X_c,Y_c)$ derived from the first step of the fit applied on the stellar components. Then we fitted the PA and the rotational velocity as for the stars. For the Seyfert 1 galaxies and for NGC\,1068, the central few arcseconds were excluded since they are highly contaminated with emission from the active nucleus (BLR or NLR).

The results of this analysis are presented in Figs. \ref{plot_pas1} and \ref{plot_pas2}.  The stellar and gaseous kinematic PAs are plotted as a function of radius. The values of the global stellar and gaseous kinematic PAs derived in Sec. \ref{sec:gpa} are indicated on each plot as a dotted red and dashed blue lines, respectively. All angles are given relative to the photometric major-axis PA.  These plots reveal the change of orientation of the stellar and gaseous kinematic major-axis orientation as a function of radius. The global values of the kinematic PAs derived in the Sec \ref{sec:gpa} correspond only to a small part of the FOV on which the kinematic orientation is globally constant. For NGC\,3627, the global stellar and gaseous kinematic PAs derived in the previous Section correspond to values outside $R \gtrsim 15 \arcsec$ (Fig. \ref{plot_pas1}), and for NGC\,5806, our first method is not sensitive to the variations of PAs between 10 and 15$\arcsec$ (Fig. \ref{plot_pas1}). In the case of NGC\,2655, the kinematic PA of the ionised gas rises abruptly for $r<10\arcsec$ from the value found by our global method to a value close to the stellar kinematic PA. 

\begin{figure*}
\centering
\includegraphics[width=\textwidth]{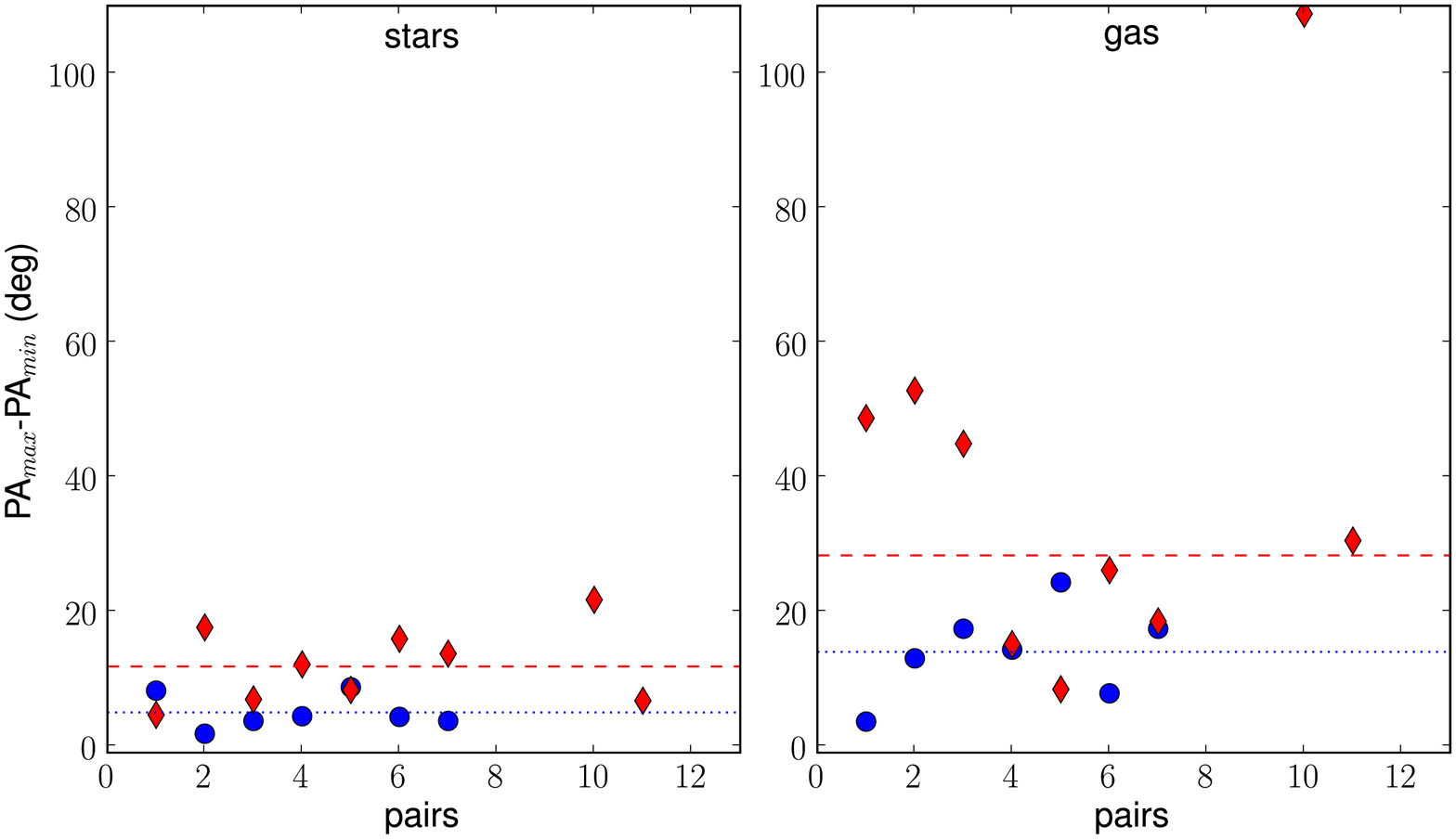}
\caption{Distribution of the amplitude of variations of the stellar (left panel) and the gaseous (right panel) kinematic PAs in the inner 1.5 kpc (see text for details). Active galaxies and inactive galaxies are shown as filled red diamonds and filled blue circles respectively. The dotted blue and dashed red lines correspond to the average of PA$_{max}$-PA$_{min}$  for the inactive and Seyfert galaxies respectively, for the stellar (left panel) or gaseous (right panel) components. NGC\,1068 and NGC\,3227 have been excluded to compute the average values corresponding to the Seyfert galaxies, since they do not have an associated inactive galaxy.}
\label{amppa}
\end{figure*}

In order to quantify the variations in the radial profiles of the kinematic PAs, we have used the difference between the maximum and minimum values of the kinematic PAs. For each galaxy, the quantity $\Delta$PA=PA$_{max}$-PA$_{min}$ is computed for the gaseous and stellar components. To compare the Seyfert hosts to their associated inactive galaxies, the maximum and minimum values of the kinematic PAs were derived on regions corresponding to the same spatial length for the two galaxies in each pair. If the velocity fields of the Seyfert and its associated inactive galaxies extend outside $r=1.5$~kpc, $\Delta$PA is derived inside this radius. If one or both of the paired galaxies extend less that 1.5 kpc, the minimum radius value of the two galaxies is used. The central few arcseconds where residual AGN-induced perturbations may be present are excluded.  The resulting $\Delta$PA values for both the gaseous and stellar components, and for all galaxies in our sample (Seyfert and inactive) are provided in Fig.~\ref{amppa}. Also shown is the {\em average} $\Delta$PA of each component for the inactive galaxies (blue lines) and the Seyfert galaxies (red lines). For the latter, the average is made excluding NGC\,1068 and NGC\,3227 since these two Seyfert do not have associated control galaxies. 

\begin{subfigures}
\begin{figure*}
\centering
\includegraphics[height=20cm]{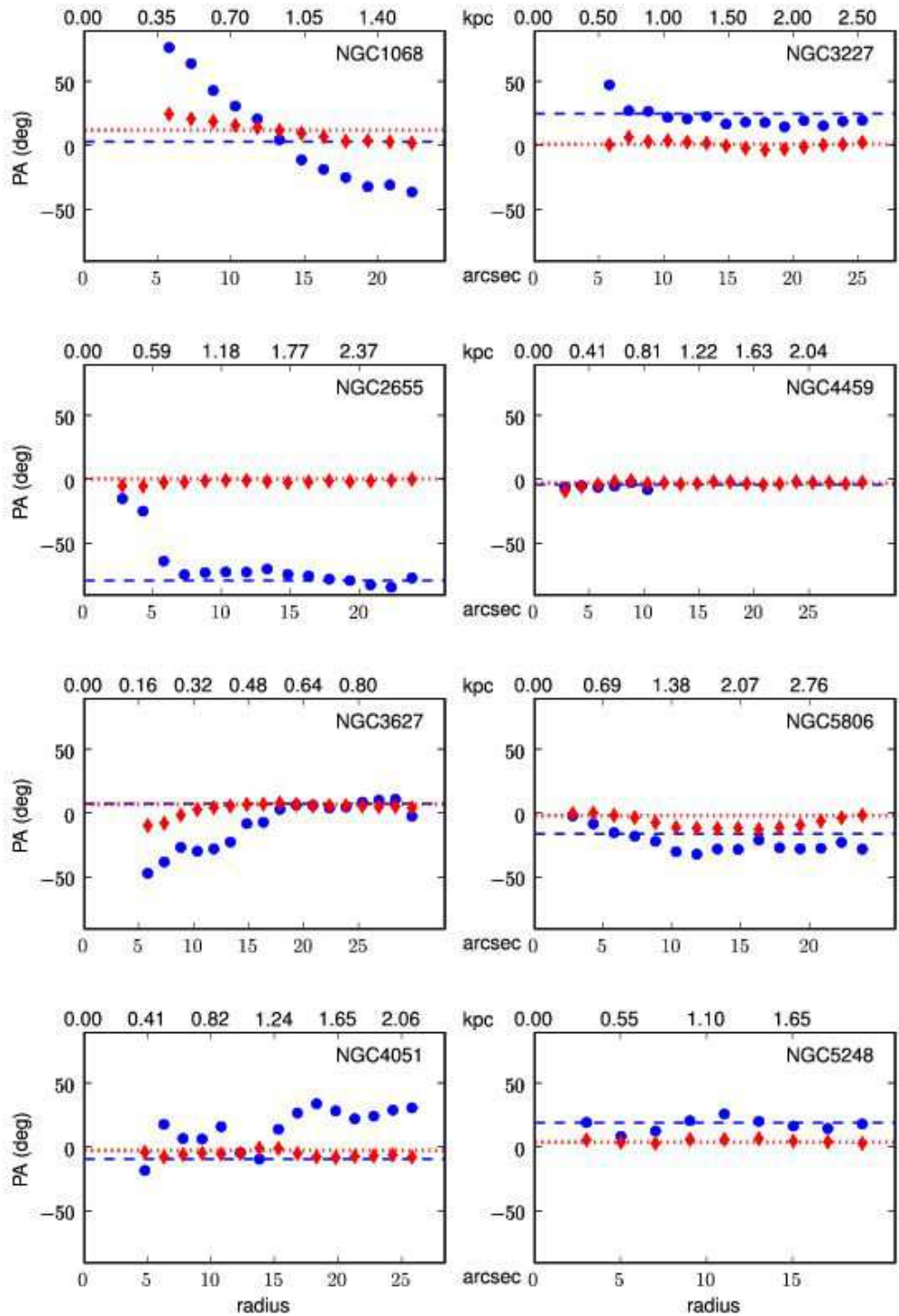}
\caption{Kinematic position angles of the stars (red diamonds) and of the gas (blue circles) components as a function of radius. The first row shows the Seyfert galaxies NGC\,1068, and NGC\,3227. Each subsequent row presents a pair of galaxies: the Seyfert on the left, and the inactive galaxy on the right. The regions related to the BLR of the Seyfert 1 galaxies (NGC\,3227, NGC\,4051 and NGC\,4151) and to the NLR of NGC\,1068 have been excluded from the fit. The dotted red and dashed blue horizontal lines represent the values of the global position angle of the kinematic major-axis for the stars and gas respectively. These values were derived from the velocity field symmetrization method (see Section \ref{sec:gpa}). The values are relative to the photometric PA  (see Table \ref{table:pas}). }
\label{plot_pas1}
\end{figure*}
\begin{figure*}
\centering
\includegraphics[height=22cm]{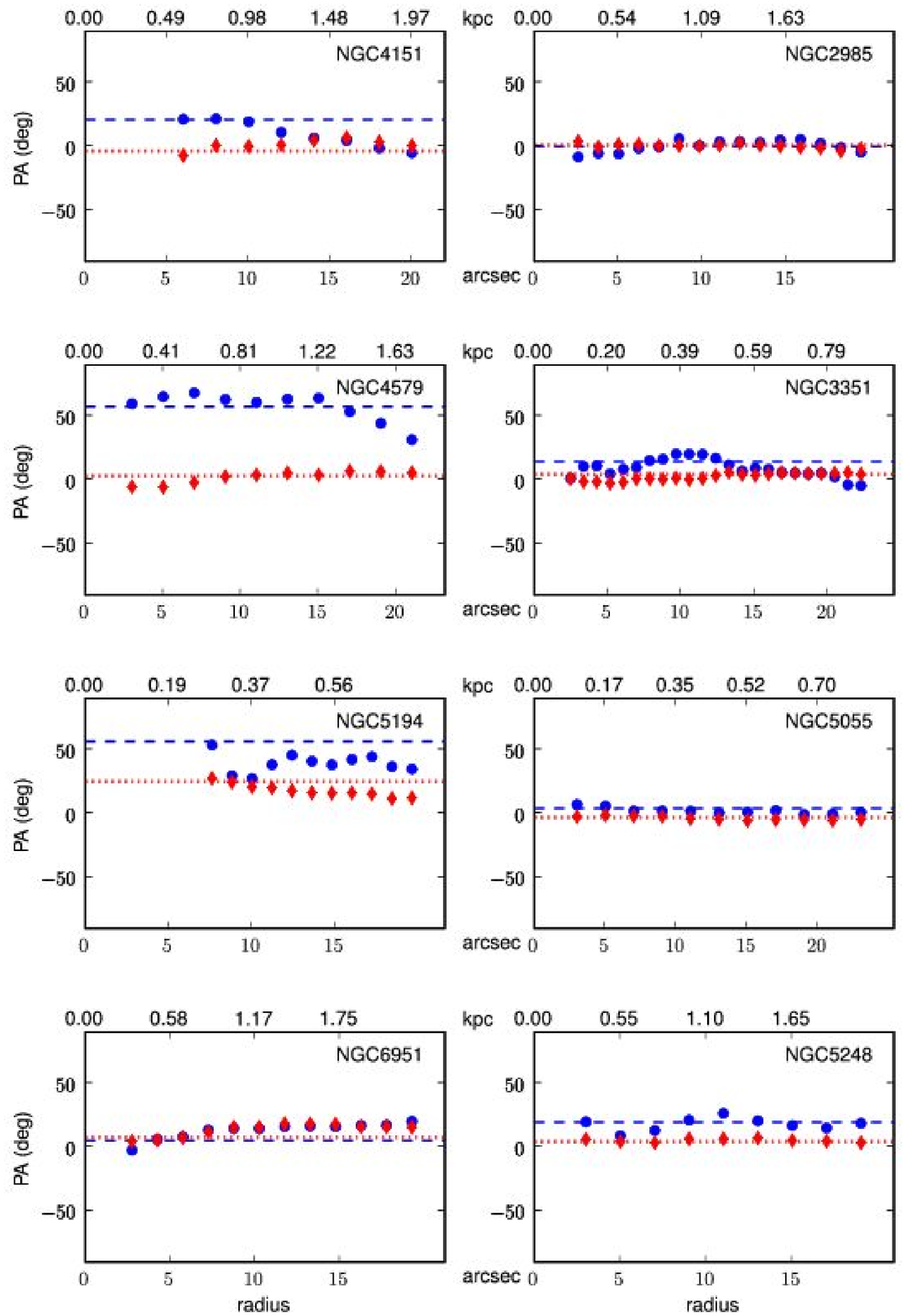}
\caption{ Fig. \ref{plot_pas1} continued.}
\label{plot_pas2}
\end{figure*}
\end{subfigures}

The first result of this analysis comes from a comparison between the gaseous and stellar components: for both the Seyfert and inactive sub-samples, $\Delta$PA is almost always higher for the gas than for the stars. While the stellar kinematic PAs radial variations are restrained within about 10$\degr$ and 20$\degr$ for the inactive and Seyfert galaxies, respectively, such variations of the gaseous PAs reach more than 40$\degr$ in the case of four Seyfert galaxies (NGC\,1068, NGC\,2655 and NGC\,3627 and NGC\,4051). The average $\Delta$PA of the gaseous component is about $\sim 3$ times greater than the average $\Delta$PA of the stars for the Seyfert galaxies as well as for the inactive galaxies.  The second result is obtained when comparing the Seyfert and inactive galaxies: for both gas and stars, $\Delta$PA is almost always larger for the Seyferts than for their control inactive galaxies. The average $\Delta$PA for the Seyferts  is about 2.5 times greater than the average $\Delta$PA for the inactive galaxies, considering either the stellar or the ionised gas components.

%% file: discussion_conclusion.tex
\section{Discussion and Conclusion}
\label{sec:discussion_conclusion}

\subsection{Accretion Rates and Kinematic Misalignments}
\label{sec:discussion1}

\begin{figure}
\centering
\includegraphics[width=8.4cm]{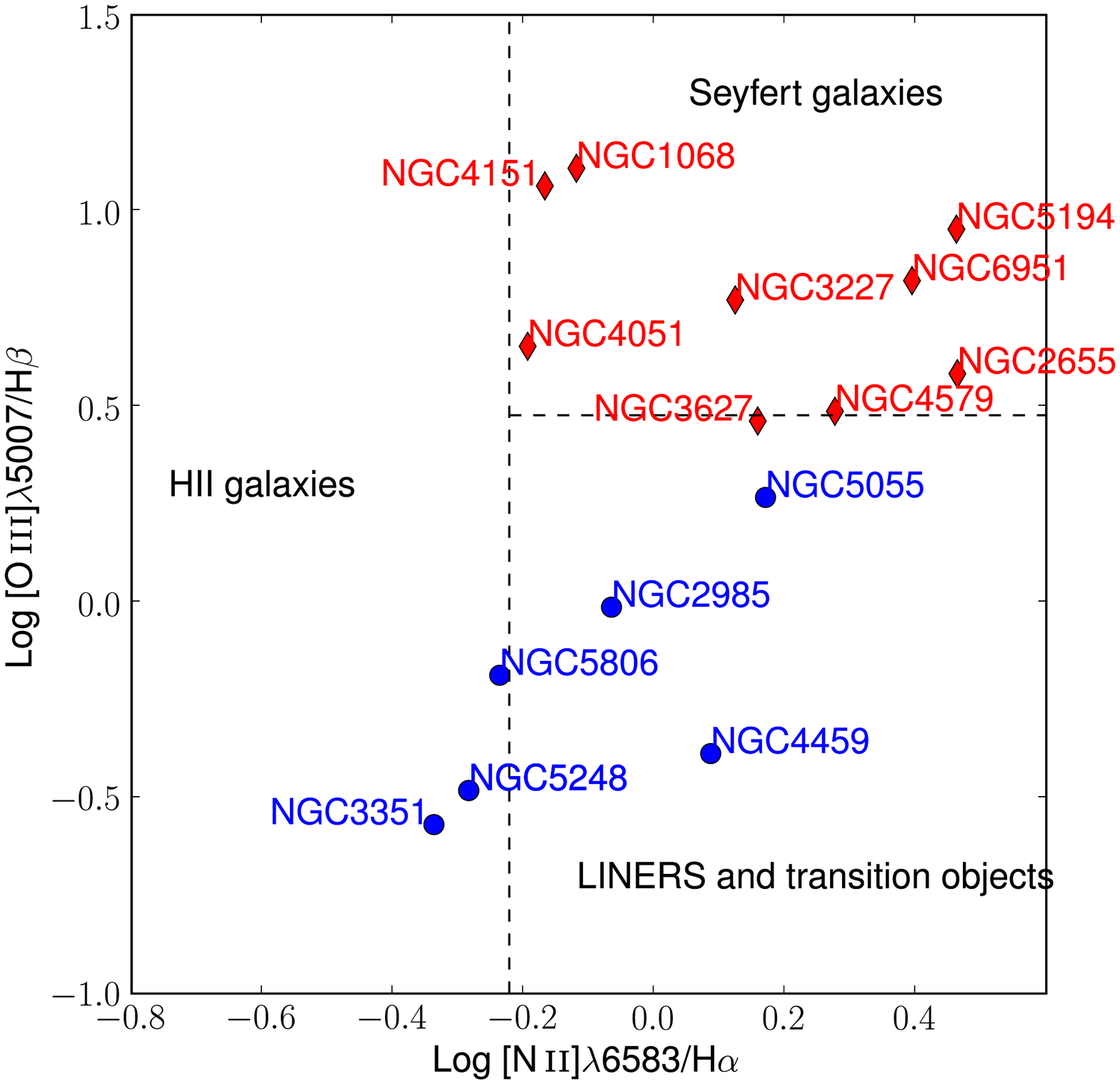}
\caption{Diagnostic diagram for the galaxies of our sample, Log(\OIIIwb/\Hb) vs Log(\NIIwb/\Ha). The emission line ratios are from Ho et al. (1997). Active galaxies are shown as filled red diamonds and inactive galaxies as filled blue circles. Finally, the classification boundaries from Ho et al. (1997) are shown as black dashed lines.}
\label{fig:diagnostic}
\end{figure}

In this section we highlight the relationship between kinematic
  disturbances in the galaxy and the accretion rate of the
  nuclear black hole.
Throughout this paper we use the nuclear type classification of \cite{Ho3}. Figure~\ref{fig:diagnostic} presents our galaxies in the diagnostic diagram Log(\OIIIwb/\Hb) vs Log(\NIIwb/\Ha), as well as the boundaries between the different excitation classes adopted by \cite{Ho3}. The separation between Seyfert and LINERS objects is not as sharp as depicted by this diagnostic diagram. We can notice that NGC\,5055 is close to be a Seyfert while NGC\,3627 and NGC\,4579 are Seyferts and close to the LINERS and Transition objects.  The activity classification of the sample nuclei therefore does not necessarily accurately represent the nuclear black hole accretion rate. We have then estimated the Eddington rates and the corresponding mass accretion rates of the individual nuclei from four relatively independent methods. The results are summarized in Table~\ref{table:Ledd}. The Black Hole masses in this Table (column~3) are from direct dynamic measurements (e.g. NGC\,1068), from reverberation mapping results (e.g. NGC\,4051, NGC\,4151), or estimated from the central stellar velocity dispersion in our SAURON maps via the well-studied M-$\sigma$ relationship \citep[e.g.][]{tremaine_2002}. We have used four observational estimates for the Eddington ratio of the nuclei. The first is the optical continuum luminosity of the galaxy nucleus from HST images (column~4 of Table\ref{table:Ledd}), interpreted as optical synchrotron radiation from the base of the jet \citep{chiaberge_2005}. No bolometric correction is applied so the result should be considered a lower limit to the Eddington ratio. Of course, contamination of the nuclear optical continuum from star formation could cause the AGN-related luminosity value to be lower. We have also used the hard (2--10 keV) X-ray luminosity, the nuclear \OIII\ luminosity (from \citealt{panessa_2006} and \citealt{Ho3})  and the nuclear radio power to estimate the bolometric luminosity of the nucleus (columns~5 to 7 of Table~\ref{table:Ledd}). In the case of the \OIII\ luminosity, we tested the estimator used for active galaxies in the SDSS   --  L$_{\rm bol}$ = 3500 $\times$  L$_{\OIII}$ \citep{heckman_2004}. This formula resulted in too high values of L$_{\rm bol}$ for our sample. The factor 3500 derived by \cite{heckman_2004} was based on the empirically derived ratios L$_{\rm 5000}$/L$_{\OIII}$ = 320, where L$_{\rm 5000}$ is the monochromatic continuum luminosity $\lambda$P$_\lambda$ at 5000\AA and L$_{\rm bol}$/L$_{5000}$ = 10.9. \cite{heckman_2004} based and tested this relationship mainly on AGN with Log \OIII\  luminosities of 6.5 and above. Our sample has much lower Log \OIII\  luminosities, ranging from 3.8 to 8.3, with median value 5.1. We find that a factor of 90, instead of 3500, gives a more reasonable value for the bolometric luminosity within our sample. This lower factor could reflect a higher \OIII\ emission relative to L$_{\rm 5000}$ from our lower luminosity, gas rich systems as compared to the SDSS Seyferts. In the case of nuclear radio luminosities, we used the sub-arcsecond AGN-related radio flux to calculate the implied jet mechanical power (see \citealt{neil_2005} for details). \cite{neil_2005} have demonstrated that this method to calculate the bolometric luminosity works well in low
  luminosity AGN. For each nucleus, we then adopted a value  of l$_{\Edd} =$ L$_{\rm bol}$/L$_{\rm Edd}$ (column~8 of Table~\ref{table:Ledd}) from the previous four values, keeping in mind their relative uncertainties and the paradigm that low luminosity AGNs in the low (high) state are more likely to produce radio (X-ray) emission. This adopted value of the Eddington ratio was used to calculate the estimated mass accretion rate (column~9). The `inactive' galaxies NGC\,3351, NGC\,5248, and NGC\,5806, all three with HII-type nuclei, have not been detected in the hard X-ray nor radio. Their Eddington ratios are likely lower than the value adopted in Table~\ref{table:Ledd}, since a significant part of their nuclear \OIII\ luminosity could come from HII regions. Note that while the overall `Seyfert' and `inactive' samples show expectedly different Eddington ratios, there is some overlap; two of the Seyferts, NGC\,3627 and NGC\,6951, have Eddington ratios similar to the inactive nuclei. 

\begin{figure*}
\centering
\includegraphics[width=\textwidth]{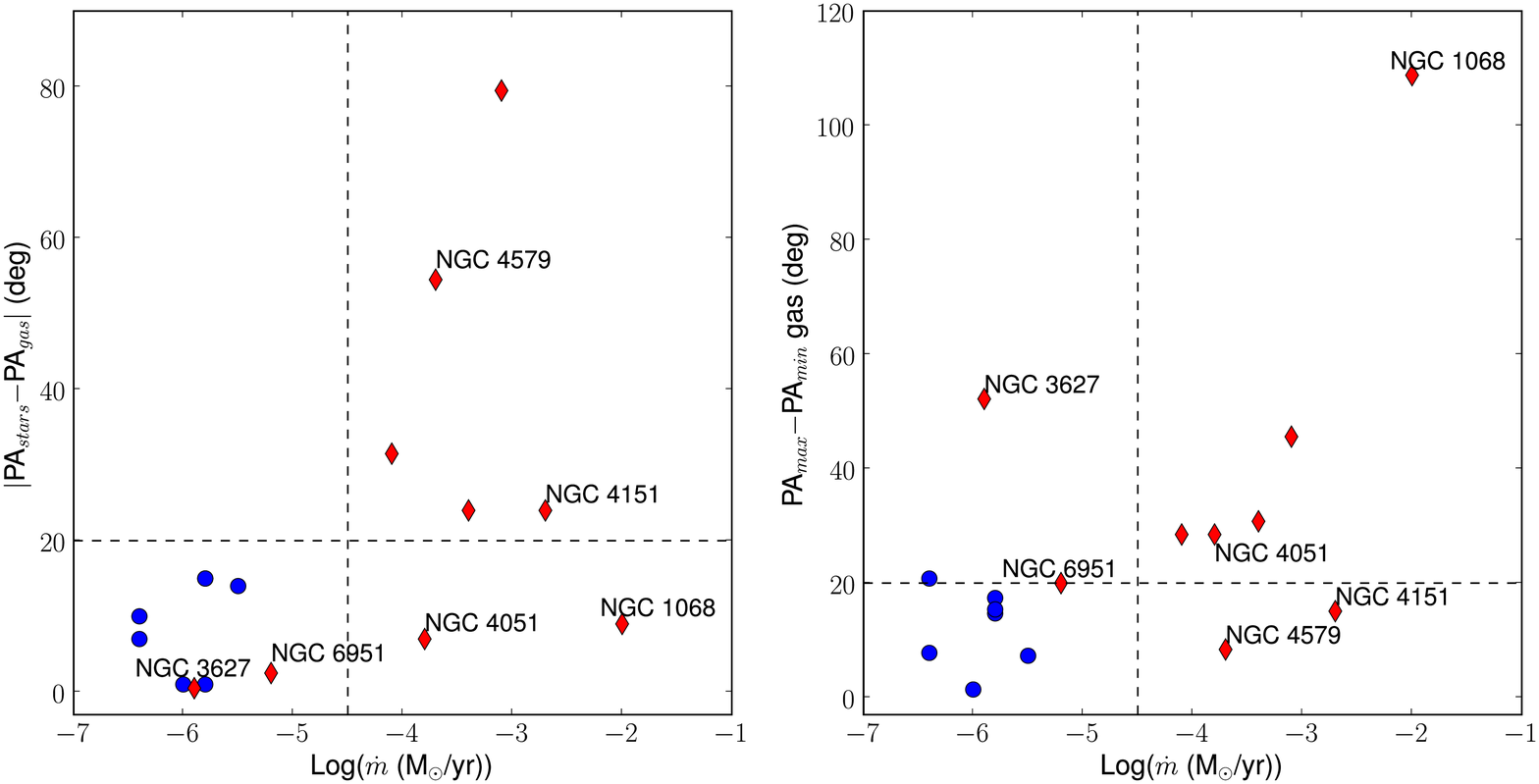}
\caption{Distribution of the differences between
the global kinematic PAs of the stars and gas (left panel) and the amplitude of variations of the gaseous kinematic
PAs in the inner 1.5kpc (right panel)
as a function of the mass accretion rate (see Table \ref{table:Ledd}). Active galaxies
are shown as filled red diamonds and inactive galaxies as filled blue circles.
For reference, a line is drawn at $\Delta$PA=20$\degr$ and mass accretion rate = 10$^{-4.5}$ M$_{\sun}$/yr in each of the panels.\label{fig:pavsLedd} }
\end{figure*}

  The relationship between mass accretion rate and the kinematic of the gas and stars is illustrated in Fig.~\ref{fig:pavsLedd}. The result is rather striking: black holes with low accretion rates reside in galaxies in which the stellar and gas kinematic axes are closely aligned and in which the gas kinematic axis is least disturbed, while Black holes with accretion rates higher than 10$^{-4.5}$ M$_{\sun}$/yr reside in galaxies with either either a large misalignment between gas and stellar axes, or a large twist in their gas kinematic axes. 

\begin{table*}
\centering
\begin{tabular}[h]{|l|l|l|l|l|c|c|c|c|}
\hline
Pairs	&Name	& Log M$_\bullet$ & Log L$_{\rm opt}$/L$_{\rm Edd}$ & Log L$_{\rm bol}$/L$_{\rm Edd}$ & Log L$_{\rm bol}$/L$_{\rm Edd}$ & Log L$_{\rm bol}$/L$_{\rm Edd}$ & Log L$_{\rm bol}$/L$_{\rm Edd}$ & Log $\dot{m}$\\
	&NGC	&(M$_{\sun}$)	&(HST)	&(X-ray)	&(\OIII)	&(Radio)	&(Adopted)	&(M$_{\sun}$/yr)\\
(1)	&(2) &(3) &(4)    &(5)    &(6) &(7)    &(8) &(9)\\
\hline
	&1068&6.90&\nodata&-1.3   &-1.1&-1.6   &-1.3&-2.0\\
\cline{2-9}
	&3227&7.59&-3.8   &-3.1   &-3.2&-3.5   &-3.3&-3.4\\
\hline
1	&2655&8.14&\nodata&-3.6   &-4.4&-3.8   &-3.6&-3.1\\
	&4459&7.81&\nodata&-6.2   &-6.1&$<$-4.3&-6.2&-6.0\\
\hline
2	&3627&7.22&\nodata&$<$-6.6&-4.6&-4.3   &-5.5&-5.9\\
	&5806&7.07&\nodata&\nodata&-4.9&\nodata&-4.9&-5.5\\
\hline
3	&4051&6.11&-2.8   &-2.1   &-2.4&$<$-2.4&-2.3&-3.8\\
	&5248&7.27&\nodata&\nodata&-5.4&\nodata&-5.4&-5.8\\
\hline
4	&4151&7.18&-2.1   &-2.0   &-1.9&-2.6   &-2.2&-2.7\\
	&2985&7.76&\nodata&\nodata&-5.9&$<$-4.1&-5.9&-5.8\\
\hline
5	&4579&7.79&\nodata&-4.0   &-4.5&-3.2   &-3.8&-3.7\\
	&3351&7.13&\nodata&\nodata&-5.9&\nodata&-5.9&-6.4\\
\hline
6	&5194&6.34&\nodata&-2.7   &-2.6&$<$-3.1&-2.8&-4.1\\
	&5055&7.06&\nodata&-6.0   &-5.1&$<$-4.0&-5.8&-6.4\\
\hline
7	&6951&7.47&\nodata&-5.1   &-4.9&$<$-3.6&-5.0&-5.2\\
	&5248&7.27&\nodata&\nodata&-5.4&\nodata&-5.4&-5.8\\
\hline
\end{tabular}
\caption{Eddington Ratios and Mass Accretion Rates. Columns are: (1)~Pair; (2)~Galaxy name; (3)~Black Hole mass from   \citet{lodato_2003} (NGC\,1068, maser dynamics),    \citet{kaspi_2000} (NGC\,3227, NGC\,4051, NGC\,4151; reverberation mapping),    \citet{Sarzi_2001} (NGC\,4459). For the other galaxies, we estimated the    black hole mass from the average stellar $\sigma$ in the central    10{\arcsec} of our \Sauron\ maps using the \citet{tremaine_2002} relationship; (4)~ratio of nuclear (HST) optical luminosity (with no bolometric correction    applied; from \citealt{chiaberge_2005}) to Eddington luminosity; (5)~Eddington ratio derived from the 2-10~keV X-ray luminosity, using     the emperical relation found by \citet{ulvestad_2001}:     L$_{\rm bol}$ = 6.7 L$_{\rm X}$(2-10~keV). The data are from the compilation    in \citet{panessa_2006}, except for NGC\,4459 and NGC\,5055 \citep{gonzalez_2006} and NGC\,6951 (\citealt{fabbiano_1992}; from $Einstein$ data  with no correction applied to transform the luminosity to the 2--10 keV regime); (6)~Eddington ratio derived from the \OIII\ luminosity using    L$_{\rm bol}$ = 90 * L$_{\OIII}$ (see text); (7)~Eddington ratio implied by the mechanical energy in the radio jet    (Q$_{\rm jet}$) as derived from the nuclear radio emission (see    \citealt{neil_2005}); (8)~adopted Eddington ratio: this was roughly calculated from columns 3 to 6     keeping in mind the reliability of the various values; (9)~the mass accretion rate implied by the adopted Eddington ratio under the standard assumption of a 10\% radiative efficiency for the accretion process.  \label{table:Ledd} }
\end{table*}

\subsection{Feeding the AGN}
\label{sec:discussion2}

Why a fraction of nearby galaxies host an active nucleus still remains an unsolved question. Several mechanisms may be involved in the fueling of the central regions of galaxies, and subsequently activating and sustaining nuclear activity. Removal of angular momentum and transportation of the gas towards the very central region may be driven by gravitational perturbations due to interactions such as mergers of galaxies or tidal interaction \citep{Barnes_1991,Barnes_1996}, or non-axisymmetric galactic structures \citep[e.g. bars, spirals,]{shlosman_2000}, as well as turbulence instabilities and shocks in the interstellar medium. The majority of optical and near infra-red (NIR) imaging studies failed to find any morphological difference between active and inactive galaxies on a range of scales that encompasses galactic interactions, stellar bars and nuclear spirals \citep{2003_martiniII,Knapen_2000,Marquez_2000}. However, \citet{Hunt04} found significant excess of isophotal twists and morphological disturbances in Seyfert 2 galaxies compared to inactive and Seyfert 1 or LINER galaxies. Another recent study of the presence and structures of dust in the circumnuclear regions revealed differences between early type Seyfert and inactive galaxies \citep{2007_dust_martini}.  These results suggest the possibility of identifiable dynamical differences, in the circumnuclear regions in active and inactive galaxies. 

In this study, we investigated the stellar and ionised gas kinematics of a well-defined sample of nearby active and inactive galaxies, in order to probe the galactic potential in the central kpc regions.  First, we have measured the kinematic misalignments between the stellar and the ionised components. The distribution of kinematic misalignments has often been used to constrain the origin of the ionised gas in early-type galaxies \citep{bertola_92,paper5}. If the gas is produced internally (e.g. stellar mass loss) and comes from the galactic disc, the bulk of the gaseous component in the central region is expected to co-rotate with the stars. The angular momenta of the stellar and gaseous components would therefore be globally aligned. Some mild misalignments may exist due to the effect of non-axisymmetric structures (bars, spirals), or the presence of decoupled stellar components. Gas acquired from an external source, e.g. via interaction with galaxies or the intergalactic medium, should eventually settle in the existing principal planes of the gravitational potential. In a triaxial potential hosting a large-scale disc, the gas would thus preferentially fall onto either the disc equatorial plane or in a plane perpendicular to it, giving rise to a polar ring. Polar rings are observed in a few nearby early-type galaxies \citep[see e.g.][]{bertola_92}. The distribution of kinematic misalignments, measured as position angle differences between kinematic major-axes, around the reference values of 0, 90 and 180$\degr$, should therefore be closely linked with the origin of the accreted gas.

 So far, there seems to be no trend for the kinematic misalignments of early-type galaxies with respect to Hubble type, galactic environment and galaxy luminosity \citep{paper5}. In our sample, five Seyfert galaxies present significant kinematic misalignments ($>$20$\degr$) however, only NGC\,2655 presents a gaseous component rotating almost perpendicularly to the stellar component. For all the other Seyfert (four objects) and for all the inactive galaxies, the ionised gas co-rotates with the stars (kinematic misalignments smaller than 20$\degr$). This follows the results of \cite{paper5} for a sample of later-type galaxies and therefore suggests that the presently observed ionised gas in the circumnuclear regions does not have a purely external origin for both the Seyfert and the inactive galaxies. The only clear exception is NGC\,2655, a galaxy known to host a polar ring \citep{2655_2}. It is interacting with a companion and had presumably suffered a minor merger in the past \citep{2655_interaction}. Interaction between galaxies is an efficient way to transport gas towards the central parts and to trigger nuclear star formation \citep{Barnes_1991}, but it seems neither a necessary nor a sufficient mechanism for most Seyfert activity \citep{derobertis_1998,schmitt_2001,ho6}.

A more refined analysis revealed that the radial variations of the kinematic PAs of the ionised gas are three times greater for the Seyfert than for their associated inactive galaxies, in the central 1.5 kpc. Indeed, due to the relatively large pixel size of SAURON compared with other IFUs and the conservative approach of our analysis, the intrinsic gaseous deviations may be even larger than suggested by our results, particularly for the very inner regions. Although our sample is small, this trend hints for the presence of non-axisymmetric perturbations of the gravitational potential in the inner kpc of the Seyfert galaxies. Since the observed stellar velocity fields of these active galaxies are very regular with major axes rather well aligned with the outer photometric ones, theses dynamical perturbations must be small. The gas, being more sensitive to small deviations from axisymmetry, responds better to small non-axisymmetric structures such as weak bars or spirals. This would be in fact in agreement with previous imaging studies which found no strong excess of bars or nuclear spirals in Seyfert galaxies. If strong deviations from axial symmetric were present, the gaseous kinematics would have been extremely disturbed  and easily detectable with standard (long-slit) spectroscopy techniques.  The presence of weak deviations from axial symmetry in Seyfert galaxies seems to be in good agreement with the refined imaging study of \citet{Hunt04}. They found an excess of NIR isophotal twists in Seyfert 2 hosts compared to inactive galaxies, and discussed the non-axisymmetric instabilities (nuclear disc, nested misaligned bars, triaxial structures) which could be responsible for such morphological perturbations. This excess of isophotal twist is not observed for the Seyfert 1 galaxies, \citet{Hunt04} interpreted this difference between Seyfert 1 and Seyfert 2 galaxies as an evolutionary effect, the Seyfert 1 hosts being 'older' than the Seyfert 2 ones in their scenario. We would need a larger sample  to investigate statistical differences in the ionised gas properties of Seyfert 1 and 2 galaxies. A larger study of Active vs. Inactive galaxies in the SDSS with IMACS is now in progress within our group, the active and inactive galaxies  drawn from the same parent sample so hopefully reducing any underlying selection biases and also offering more controls per Seyfert (Westoby et al. in preparation).

 In addition to a ready supply of fuel and a transportation mechanism, timescales are important. Galaxies are thought to experience recurrent episodes of nuclear activity in their lifetime, with instabilities in the fueling rate possibly regulating the activity cycles \citep{saripalli_2007astro}. Dynamical instabilities in the inner kpc regions and nuclear activity can evolve in parallel. However, due to the time required to set up sufficient gas inflow, the non-axisymmetric  perturbations may be observed while not enough gas has been driven in the vicinity of the SMBH and the galaxy is inactive. Similarly such gravitational instabilities may have disappeared when the nuclear activity is still underway.  Observationally, it is therefore reasonable to hypothesise that a snapshot of active and inactive galaxies might show equal degrees of host galaxy disturbance.  However, while disturbed gas observed in a host galaxy is not the gas that is simultaneously fueling the nuclear activity, the dynamical time scale in the circumnuclear regions is similar to the AGN lifetime. Therefore, the correlation between the presence of perturbation in the ionised gas in the central kpc regions and the nuclear activity suggested by our study is striking and supports a close link between the AGN, host dynamics on scales of hundreds to tens of parcsecs and relevant dynamical and activity timescales. 

Finally, the presence of $\sigma$-drops in the stellar velocity fields  observed for some of our galaxies may indicate the presence of cold nuclear stellar discs \citep{eric_2001}. \cite{wozniak_2003} interpreted such structures as the result of star formation following the accretion of gas in the central few hundred parsecs. High resolution follow-up IFU mapping of the central few arcseconds, using e.g. the OASIS IFU, could reveal direct evidence for these discs. Such data on a relatively small FOV are rather difficult to interpret. However, associated with our larger scale SAURON maps, they could  extend our understanding of fuel delivery closer again to the AGN.

\subsection{Summary}
\label{sec:conclusion}

In this paper, we have presented two dimensional morphology and kinematic maps of the stars and ionised gas component in the central kpc regions of a distance limited sample of well-matched Seyfert and non-Seyfert galaxies. We found that: 
\begin{itemize}
\item Kinematics of the stellar and gaseous components in the circumnuclear regions of our galaxies are dominated by disc-like rotation. The stellar kinematics present very regular rotation patterns for both inactive and Seyfert galaxies in agreement with previous studies \citep{barbosa_2006}. Ionised gas is generally co-rotating with respect to the stars.  
\item The stellar global kinematic major-axes are aligned within 10$\degr$ with the outer galactic disc major-axis for all galaxies, except for two Seyferts:  NGC\,1068 and NGC\,5194, for which the determination of the photometric major axis orientation is quite uncertain. Except for these two particular galaxies, the circumnuclear stellar kinematic major axis is a robust determination of the line-of-nodes for either the Seyfert and the non-active galaxies.
\item The gaseous component co-rotates with the stars, the global stellar and gaseous kinematic PAs being aligned for all galaxies within 20$\degr$, except for two Seyfert galaxies:  NGC\,2655 and NGC\,4579. Our kinematic maps reveal some disturbances in the gaseous velocity field, which may be evidence for deviations from axisymmetry. These perturbations in the ionised velocity fields may hint for the presence of streaming or radial flows, which could be related to fueling mechanisms.

\item The gaseous kinematics PAs have radial variations over the SAURON FOV 3 times greater than the stellar ones, for either the Seyfert and their associated inactive galaxies.
\item The variations of the stellar and gaseous PAs in the central kpc of the Seyfert galaxies are 2.5 times greater than those of the inactive galaxies.
\end{itemize}

We conclude that the increased incidence of disturbed ionised gas in
the circumnuclear regions of the Seyferts on scales of hundreds to
tens of parsecs, compared with inactive galaxies, supports a close
link between nuclear host dynamics and relevant dynamical and activity
timescales.

\section*{Acknowledgments}
We thank Michele Cappellari for the IDL implementation of the position angles routine. We would  like to warmly thank Pierre Ferruit and Bruno Jungwiert for their contribution and fruitful discussions at an early stage of this project, and we thank Tim de Zeeuw and the referee, Jack Gallimore, for useful comments that improved the
paper. CGM acknowledges financial support from the Royal Society. The \Sauron-related projects are made possible through grants 614.13.003, 781.74.203, 614.000.301 and 614.031.015 from NWO and financial contributions from the Institut National des Sciences de l'Univers, the Universit\'e Claude Bernard de Lyon I, the Universities of Durham, Leiden and the Netherlands Research School for Astronomy NOVA. The \Sauron\ observations were obtained at the William Herschel Telescope, operated by the Isaac Newton Group in the Spanish Observatorio del Roque de los Muchachos of the Instituto de Astof{\'i}sica de Canarias. We would also like to thank the WHT staff for their technical support  at the telescope. This work made use of the HyperLeda and the NED databases. Part of this work is based on data obtained from the STSci Science Archive Facility. The Digitized Sky Surveys were produced at the Space Telescope Science Institute under US Government grant NAG W-2166. The images of these surveys are based on photographic data obtained using the Oschin Schmidt Telescope on Palomar Mountain and the UK Schmidt Telescope

%% file: appendices.tex
\appendix

\section{Notes on individual galaxies}
\label{app:gal_indiv}

\subsection{NGC\,1068 (M\,77)}
NGC\,1068 is one of the most nearby Seyfert~2 galaxies ($D=14.4$~Mpc), and it has been well studied at different wavelengths. \cite{Ho3} classified this active galaxy as a Seyfert of Type 1.8, but a dusty torus hides a Seyfert~1 nucleus \citep{1068_antonucci, 1068_jourdain}. NGC\,1068's nucleus hosts a radio jet and an ionised cone at PA $\sim 10\degr$ \citep{1068_jet}. \cite{eva_1068} summarized the different observed morphological structures : the outer disk and the outer oval, the latter being interpreted as a primary bar, a HI ring at its Outer Lindblad Resonance and the two-arms inner spiral at its Inner Lindblad Resonance. A secondary Near-Infrared (NIR) bar extends up to about 16\arcsec. The \Sauron\  data we presented here have been already published by \cite{1068_1}, and we refer the reader to this paper for a detailed analysis: we here only provide a brief description of the revealed structures.

The stellar continuum map (Fig.~\ref{pair0}) shows elliptical isophotes, elongated along the PA of the NIR bar. The stellar velocity field (Fig.~\ref{pair0}) shows strong departures from axisymmetry, with an S-shaped zero velocity line, and a slightly varying orientation of the kinematics major-axis. The velocity dispersion rises towards the centre, reaching 200 km$\,$s$^{-1}$ at R$\sim$10\arcsec and then presents a drop in the inner 5\arcsec with values down to about 100 km$\,$s$^{-1}$.  \cite{gerssen_2006} observed this galaxy with the Gemini Multi-Object Spectrograph (GMOS) IFU, covering the central 10 $\times$ 8 arcsec. Our stellar kinematics properties are in good agreement with theirs although they found a kinematic PA offset by about 13$\degr$ from ours, their FOV being too small to detect the change of orientation of the kinematic major-axis.

~\Hb\ and \OIII\  distributions are quite different (Fig. \ref{pair0}). \Hb\ emission is very high in the inner 5\arcsec, and traces the spiral arms outside this region. As for \Hb\ , \OIII\  emission peaks in the central parts, with the distribution of \OIII\ becoming very asymmetric outwards. It is found predominantly in the North-East side, tracing the northern ionisation cone. Despite the significant differences between the distribution of the \Hb\ and \OIII line emission, their velocity fields and velocity dispersion maps are very similar (Figs.~\ref{pair0} and \ref{kinHb1}). The velocity fields of \Hb\ and \OIII\  both display a prominent S-shaped zero velocity curve, evidence for strong deviations from circular motions. The velocity dispersion maps of \Hb\ and \OIII\  show a peak in the central 5\arcsec\ parts and then reach lower values outside. As expected, the \OIII/\Hb\ is high in the region of ionisation cone ($\sim$ 10), and lower in the spiral arms (\OIII/\Hb\ $<$ 1), the latter being dominated by star formation regions (Figs. \ref{pair0}).  

\subsection{NGC\,2655}
NGC\,2655 is an early-type spiral galaxy, hosting a Seyfert~2 nucleus. Optical and radio spectroscopic data exhibit off-plane gas in the central part \citep{2655_1}, and a dusty polar ring seen in \Ha\ \citep{2655_2}. This galaxy is interacting with its companion NGC\,2715 and presents remnant signatures of a past merger \citep{2655_interaction}. 

The stellar continuum map (Fig.~\ref{pair1}) shows an elongated feature, along the East-West direction, which corresponds well with the PA of the line of maximum stellar velocity. The stellar velocity field (Fig.~\ref{pair1}) is remarkably regular, with a PA around $90\degr$. The velocity dispersion map shows a drop in the inner 3\arcsec.  

~\Hb\  and \OIII\  emission lines distribution and kinematics are similar (Figs.~\ref{pair1} and \ref{kinHb1}). The emission lines are very bright in the inner 5\arcsec. Away from the centre, the distribution maps show a bright knot to the South-East side, 15\arcsec\ away from the nucleus, and a lane at 10\arcsec on the West side: both features have been reported by \cite{2655_2}. The SE knot corresponds to high \OIII/\Hb\ ratio values ($\sim 4$) similar to the ones obsered in the nuclear region. This feature could be driven by the radio jet \citep{2655_2}. The emission lines ratio is smaller in the West lane (around 2-3). The emission lines velocity fields show strong departures from axisymmetry, with an S-shaped zero velocity curve, the kinematic major-axis having a changing PA, from $\sim 90\degr$ in the inner 5\arcsec\ to near $180\degr$ in the outer parts. As hinted by \citet{2655_4}, the ionised gas rotates together with the stars in the inner 5\arcsec, but follows the dust polar ring at radii $>5 \arcsec\ - 6 \arcsec$, rotating perpendicularly to the galactic plane. \OIII\  and \Hb\ velocity dispersion maps show the same features: a rise in the inner 5\arcsec\ ($\sim 220$~km$\,$s$^{-1}$) and in the western feature (305~km$\,$s$^{-1}$), and a roughly constant value outside (between 100 and 150~km$\,$s$^{-1}$). 

\subsection{NGC\,2985}
NGC\,2985 is an early-type spiral galaxy, with regular gas and stellar distributions. It is the control for NGC\,4151. The disk orientation is determined by an inclination of $i=40\degr$ and a major-axis position angle of $PA=180\degr$. \cite{2985_1} report a tidal interaction with its companion NGC\,3027. 

The stellar velocity field of NGC\,2985 shows a regular and symmetric pattern, consistent with motions in an axisymmetric gravitational potential (Fig.~\ref{pair4}). The PA of the kinematic major-axis (-2$\degr$) is aligned with the photometric major-axis of the outer disc.  The velocity dispersion rises inwards, with 100~km$\,$s$^{-1}$ at 15\arcsec from the centre and 150~km$\,$s$^{-1}$ in the inner 5\arcsec. 
There is very little emission from the ionised gas \Hb\ and \OIII\ in NGC\,2985 (Fig.~\ref{pair4}). The emission lines flux peaks in the inner 2\arcsec\, with very little flux outside. The \Hb\  (Fig.~\ref{kinHb2}) and \OIII\  (Fig.~\ref{pair4}) kinematics are similar: the velocity fields are regular, showing the same overall symmetry as the stellar velocity field. The gaseous velocity dispersion maps show no specific features, with values rising towards the centre (200~km$\,$s$^{-1}$ for \OIII, 160~km$\,$s$^{-1}$ for \Hb). Finally, the \OIII/\Hb\ line ratio reaches values up to 1.5-2 in the centre and decreases poutwards (for $r>10$~\arcsec).

\subsection{NGC\,3227}
NGC\,3227 is a well studied barred galaxy, interacting with the elliptical galaxy NGC\, 3326, and with an associated dwarf galaxy \citep{carole_3227_2}. The galactic disc has an inclination of $56\degr$, with a outer photometric major-axis at a position angle $PA=158\degr$ \citep{carole_3227}, coincident with the major-axis of the stellar bar. It is hosting a type 1.5 Seyfert nucleus \citep{Ho3}. The central region has been mapped in $^{12}$CO(1-0) and $^{12}$CO(2-1) by \cite{eva_3227}, who detect molecular gas very close to the nucleus ($\sim 13$~pc). The inner kpc regions host several complex features:  a radio jet (PA $\sim -10\degr$; \citet{3227_carole_95b}), an conical NLR outflow at a PA of about 15$\degr$ \citep{carole_3227}, an \Ha\ outflow at PA~$\sim 50\degr$ \citep{3227_outflowHa},  and a molecular nuclear ring \citep{eva_3227}. 
 
Our \Sauron\  stellar continuum map (Fig.~\ref{pair0}) presents an elongated structure (PA~$= 153\degr$), aligned with the stellar bar. The stellar velocity field is regular (Fig.~\ref{pair0}), the major-axis being parallel to the galactic disc orientation. Outside $r > 5$\arcsec, the stars rotate rigidly (isovelocity contours parallel to each others). The stellar velocity dispersion rises towards the centre, reaching 200~km$\,$s$^{-1}$ in the inner 5\arcsec. 

Fig.~\ref{pair0} presents the \OIII\  and the narrow component of \Hb\  emission lines. The central broad component for the \Hb\  line corresponding to nuclear emission (BLR) has been removed. \OIII\  distribution is rather asymmetric, elongated on the North side of the field and \Hb\  emission lines extends from South-East to East. The \OIII\  and \Hb\ velocity fields are similar: they show strong deviations from axisymmetry with a PA of the major-axis changing from $\sim 30\degr$ in the central regions, to about $170\degr$, close to the stellar kinematic major-axis orientation. The velocity dispersion of both \Hb\  and \OIII\  rise towards the nucleus, reaching 310~km$\,$s$^{-1}$ and 360~km$\,$s$^{-1}$, respectively in the centre. Finally, the \OIII/\Hb\ lines ratio map shows an elongated structure of high values ($>7$) along the global kinematic major-axis (North-South). 

\subsection{NGC\,3351 (M\,95)}

NGC\,3351 is a barred inactive spiral galaxy. It is the control object for NGC\,4579. NGC\,3351 hosts a large-scale stellar bar and an inner molecular bar-like feature in the centre \citep{3351_1}. The latter molecular structure is aligned with the major-axis of the galaxy (PA~$\sim 10\degr$) and perpendicular to the large scale bar. Two rings of HII regions associated with resonances due to the stellar bar exist: one inner ring with a radius of 10\arcsec\ and another at 70\arcsec\ from the centre. More recent observations in CO \citep{3351_2} show evidence of the non-circular streaming motions in the inner region. 

The \Sauron\  stellar continuum map (Fig.~\ref{pair5}) shows a ring-like structure with $6\arcsec \la R \la 10\arcsec$, which corresponds well with the HII ring at the presumed ILR of the large scale bar. The stellar velocity field is regular throughout the field of view. Stellar velocity dispersions are slightly lower (by $\sim 20$~km$\,$s$^{-1}$) inside the ring than outside.

~\OIII\ and \Hb\ distributions and kinematics do not differ much from each others. The intensity maps clearly show the known HII ring at $\sim 10\arcsec$ (Fig.~\ref{pair5}), and there is very little emission outside that structure. The gas velocity maps are regular, following the orientation observed in the stellar velocity field. The \OIII\ velocity dispersion map is almost featureless, with only a slight gradient from North-East (100 km$\,$s$^{-1}$) to South-West (140 km$\,$s$^{-1}$). There is a velocity dispersion drop ($\sim 50$~km$\,$s$^{-1}$) for \Hb\ in the ring (Fig.~\ref{kinHb2}). The \OIII/\Hb\ lines ratio map show low values in the inner regions ($\lesssim$ 0.1), as expected from star formation. \OIII/\Hb\ ratio is significantly higher outside the ring ($\sim 1$), probably mostly due to the uncertainty in the emission line flux there.

\subsection{NGC\,3627 (M\,66)}
NGC\,3627 is an SAB spiral galaxy hosting a Seyfert~2 nucleus. It is part of the Leo Triplet system, interacting with NGC\,3623 and with evidence for a past interaction with NGC\,3628 \citep{3627_interact}. Previous radio and optical observations show asymmetric structures in the velocity field, non-circular gaseous motions and a molecular CO inner ring \citep{3627_1}. More recently, IFS data obtained by \cite{3627_2} showed evidence for strong gaseous radial motions.

The \Sauron\  stellar continuum map shows an elongated structure along the photometric PA $\sim 170\degr$ (Fig.~\ref{pair2}). The velocity field presents a regular rotation pattern, with a slight twist of the zero-velocity line in the very central parts ($r < 3 \arcsec$). The stellar rotation follows the orientation of the photometric major-axis. The stellar velocity dispersion rises towards the central regions up to 120~km$\,$s$^{-1}$. 

The \OIII\ emission is concentrated in the inner 5\arcsec, with some emission in an elongated structure, along the photometric PA (Fig.~\ref{pair2}). As for \OIII, \Hb\ emission is preponderantly observed in the central parts, with an additional bright spot about 15\arcsec\ South of the nucleus. The \OIII\ and \Hb\  kinematics are quite similar (Figs.~\ref{pair2} and \ref{kinHb1}). Their velocity fields show strong deviations from axisymmetry: the orientation of the kinematic major-axis changes from the external regions where it is aligned with the stellar kinematics, toward the nucleus where the ionised gas kinematic major-axis deviates by $\sim 40\degr$ from the stellar one. \citet{3627_2} suggested that strong radial motions may exist in the central parts of this galaxy. Both \OIII\ and \Hb\  velocity dispersion maps are almost featureless, with a slight depression in the very centre. Finally, the \OIII]/\Hb\ lines ratio reaches its highest values in the inner 2\arcsec.

\subsection{NGC\,4051}

NGC\,4051 is an SAB galaxy classified as S1.2 by \citet{Ho3}. The stellar bar extends up to $\sim 50\arcsec$ along a PA $\approx 135\degr$. Radio-observations show a triple source at a PA~$\sim 73\degr$ \citep{4051_outflow}, and the \OIII\ emission line distribution in the inner 3\arcsec\ is aligned with this radio component. The narrow-line profiles of NGC\,4051 present strong blue wings. This blue-shifted component is also detected with \Sauron\  in the \Hb\ and \OIII\ lines, and has therefore been fitted (see Sect.~\ref{sec:gasreduc}). \citet{4051_outflow} found evidence for a conical outflow, at 1\farcs5 from the nucleus. 

Our \Sauron\  stellar continuum map (\ref{pair3}) show regular isophotes elongated along the PA of the large-scale bar. The stellar velocity field presents a regular and symmetric rotation pattern, the rotation axis being aligned with the photometric major-axis (PA$\sim 135\degr$). The stellar velocity dispersion decreases in the inner 5\arcsec down to $\sim 50$~km$\,$s$^{-1}$. 

The emission line intensity is slightly extended towards the North-East, consistent with the ionised \OIII\ outflow cone observed by \citet{4051_outflow}. The orientation of the ionised gas kinematic major-axis varies with radius: it is aligned with the stellar one in the outer part of the \Sauron\  FOV, and then deviates by about $30\degr$ from the line-of-nodes of the galaxy. The \Hb\ emission line velocity dispersion reaches its highest values in the inner 5$\arcsec$, while a decrease of dispersion is observed in these central regions for \OIII. The \OIII/\Hb\ lines ratio peaks at about 3$\arcsec$ North-East from the nucleus, within the outflow region \citep{4051_outflow}. 

\subsection{NGC\,4151}

NGC\,4151 is one of the most well-studied galaxies of our sample. For a review on this galaxy and the AGN properties, see \citet{4151_review}. It is an almost face-on barred spiral galaxy hosting a Seyfert type 1.5 active nucleus \citep{Ho3}. The large-scale weak bar is elongated along a PA of about 130$\degr$ and the photometric major-axis orientation is close to 20$\degr$ \citep{carole_4151_2}. Its nuclear continuum and BLR emission are highly variable. It contains a radio jet along a PA of 77$\degr$ and its Narrow Line Region (NLR) is extending over $\sim 10\arcsec$, and is aligned with the jet \citep{carole_4151_jet}.

Our stellar continuum map (Fig.~\ref{pair4}) presents rather round isophotes. The stellar velocity field shows a twist of the zero-velocity line in the inner 5 $\arcsec$ which corresponds to the region where the BLR dominates the spectral features. A $\sigma$-drop is observed in the central part of the velocity dispersion map, but could be due to the BLR contamination. The \OIII\ and \Hb\ emission lines distributions are elongated on the South-West direction at a PA of about 50$\degr$ associated with the NLR and the Extended NLR \citep[ENLR, $R > 5\arcsec$][]{4151_carole_2005}. The emission lines ratio reaches very high values ($\sim 12$) in this region. The ionised gas kinematics is dominated by rotation, but with some disturbances on the North-West side. \OIII\ emission line velocity dispersions peak in the inner few arcsec ($\sim 280$~km\,s$^{-1}$). Outside the very central parts, the velocity dispersion is between 50 and 130~km\,s$^{-1}$. It increases at the edges of our map, corresponding to the location of a dusty ring \citep{4151_dust}. \cite{4151_dustkin} revealed the presence of streaming motions along this dust structure, which may be associated with the high \OIII\ velocity dispersion observed in our \Sauron\  maps. \Hb\ velocity dispersion is flat in the outer parts ($R > 10\arcsec$, with $\sigma$ between 80 and 130~km\,s$^{-1}$) and rises regularly towards the center, reaching 200 km\,s$^{-1}$ in the inner 3$\arcsec$. The maximum velocity dispersion ($\sim 350$~km\,s$^{-1}$) is reached at about 3\farcs5 North-East of the centre. 

\subsection{NGC\,4459}

This S0 galaxy is in the Virgo cluster. It is the control non-active galaxy for NGC\,2655. The \Sauron\  data presented here (Fig.~\ref{pair1}) are part of the \Sauron\  project \citep{paper2} and were therefore already published elsewhere \citep{paper3, paper5}.

The stellar component of NGC\,4459 shows roundish isophotes, and its stellar velocity field presents a very regular rotation pattern the major-axis of which is parallel to the line-of-nodes (PA~$\sim 77\degr$). The velocity dispersion rises towards the centre, reaching 200~km$\,$s$^{-1}$ (Fig.~\ref{pair1}).
There is almost no ionised gas emission (our maps have been clipped at $3 \sigma$), except for the very centre ($R < 3\arcsec$).The emission lines velocity field seems to be dominated by rotation, and the velocity dispersion rises from about 80~km$\,$s$^{-1}$ for $R > 3\arcsec$ to 160~km$\,$s$^{-1}$ for \Hb\  (200~km$\,$s$^{-1}$ for \OIII) inside 3\arcsec. The \OIII/\Hb\ line ratio rises regularly towards the centre.

\subsection{NGC\,4579 (M\,58)}

NGC\,4579 is an active galaxy classified as S1.9/L1.9 \citep{Ho3}. It hosts a NIR stellar bar of 9~kpc of diameter, oriented along PA~$=58\degr$. \cite{nugaIV} observed this galaxy with the Plateau de Bure Interferometer (CO) as part of the NUGA survey. They found a nuclear molecular spiral from $R \sim 1$~kpc down to $\sim 200$~pc, driven by the stellar bar and detected highly non-circular motions over the spiral arms. They interpreted these perturbations as outflow motions.

Fig.~\ref{pair5} presents our \Sauron\  maps for this galaxy. The stellar continuum isophotes are rather round in the centre and slightly elongated along a PA of $\sim 60\degr$ corresponding to the orientation of the stellar bar. The stellar velocity field presents a very regular rotation pattern, with a kinematic major-axis oriented at a PA of 95\degr. The stellar velocity dispersion rises regularly towards the centre. 

~\OIII\ and \Hb\ distribution and kinematics are very similar (Figs.~\ref{pair5} and \ref{kinHb2}). The ionised gas intensity maps show a spiral-like structure which corresponds to the nuclear molecular structure observed by \cite{nugaIV}. The \OIII/\Hb\ ratio presents lower values in the spiral arms than outside. The ionised gas kinematics is complex and shows strong departures from axisymmetry. The gas kinematic major-axis ($PA=100\degr$) is almost perpendicular to the stellar bar, and varies within the \Sauron\  field of view. Despite the similarities between \Hb\ and \OIII\ distributions and velocity fields, their velocity dispersions are significantly different. The $\sigma_{[OIII]}$ map presents an elongated structure of high values along the stellar bar (Fig.~\ref{pair5}), while $\sigma_{\Hb}$ decreases in the central parts Fig.~\ref{kinHb2}.

\subsection{NGC\,5055 (M\,63)}

NGC\,5055 is a spiral non-active galaxy, control for NGC\,5194. Previous radio and optical studies show an overall remarkable regularity and symmetry in its morphology and kinematics \citep{5055_1, 5055_2}. There are evidence for a tidal interaction with a companion (UGC\,8313), and the outer disk of NGC\,5055 is warped. \cite{5055_3} studied the \Ha\ emission of NGC\,5055 with a Fabry-Perot: they found two velocity components in the inner 8\arcsec of this galaxy. One component is consistent with the global disk kinematics while the second one exhibits a pattern compatible with a bipolar outflow or a counter-rotating disk. 

Our \Sauron\  stellar continuum map presents an asymmetry towards the North-West (Fig.~\ref{pair6}). The stellar velocity field shows very regular rotation pattern with the kinematic major-axis aligned with the photometric one at a PA of 100\degr. The velocity dispersion rises in the very centre  ($r < 2\arcsec$). 

The ionised gas \Hb\ and \OIII\  are mainly concentrated in the inner 5\arcsec. The \OIII\ distribution shows a V-shaped structure towards the North, while the \Hb\ intensity map shows a number of hot spots broadly distributed over the FOV. The \OIII\ and \Hb\ velocity fields are very regular and similar to the stellar velocity field (except for their higher amplitudes). \OIII\ velocity dispersion maps is featureless but \Hb\ velocity dispersion map shows an elongated structure of high dispersion roughly aligned with the kinematic major-axis. Finally, the \OIII/\Hb\ lines ratio reaches its highest values in the region of high \OIII\ emission line flux.

\subsection{NGC\,5194 (M\,51)}

NGC\,5194 is an almost face-on spiral galaxy in interaction with NGC\,5195. Its nucleus is classified as a Seyfert type 2 \citep{Ho3}. Many studies have been made on the structure and the nucleus of this nearby galaxy. A large outflow bubble extents up to $\sim 9\arcsec$ ($\sim 42$~ pc) North of the nucleus and a bright cloud is located about 3\arcsec South of the nucleus. These structures have been observed at different wavelengths \citep{M51_Ford_bubblejet}, and they are found to trace the biconical NLR along a PA of 163\degr. NGC\,5194 is hosting a radio-jet which interacts with the surrounding interstellar medium. 

The \Sauron\  stellar continuum map (Fig.~\ref{pair5}) shows the presence of dust lanes. The stellar velocity field has a regular rotation pattern with its major-axis at a PA~$\sim -165\degr$, deviating by $\sim 25\degr$ from the presumed line of nodes orientations \citep[PA~$=170\degr$,][]{tully_M51}. The stellar velocity dispersion is featureless. 

~\Hb\ emission line flux is very high in the centre and for $R > 5\arcsec$ it traces the spiral arms of the galaxy. The \OIII\ emission line flux traces the ionisation cone of NGC\,5194. The \OIII\ intensity map shows the outflow bubble in the North and the bright cloud in the South of the nucleus corresponding to the end of the radio-jet. The emission line ration \OIII/\Hb\ peaks in the centre (with values up to 8) and in the southern cloud (values between 3 and 5). The ionised gas velocity field seems to be dominated by simple galactic rotation, but presents some distortions very probably due to the outflow observed in the inner 10\arcsec. The ionised gas velocity dispersion rises towards the centre.

\subsection{NGC\,5248}

NGC\,5248 is a nearby inactive spiral galaxy, control for both NGC\,4051 and NGC\,6951. It harbours two prominent circum-nuclear rings \citep{5248_1} and an inner molecular spiral within 1.5~kpc driven by the large scale bar \citep{5248_2}. The arms of the spiral connect with the inner circumnuclear ring. \cite{5248_2} also mention the presence of streaming motions in this galaxy. 

The stellar continuum shows an elongated structure along the global disk orientation at $PA \sim 150\degr$ (Fig.~\ref{pair3}). The velocity field presents a regular rotation pattern, the major-axis of which is oriented at a PA of 115\degr, aligned with the photometric major-axis \citep[$PA=105\deg$][]{5248_2}. A $\sigma$-drop is revealed by our velocity dispersion map in the inner 3\arcsec with an amplitude of $\sim 60$~km$\,$s$^{-1}$.  

The \Hb\ emission line distribution shows a ring-like structure with a radius of $\sim 5\arcsec$. The \OIII\ distribution is concentrated within the central few arcseconds, and in the two brightest peaks of the ring. This ring-like structure corresponds to the known ring of HII region described by \cite{5248_1}. The \OIII/\Hb\ emission line ratio is lower inside the ring (0.15), consistently with the presence of star formation, and rises slightly towards the very centre (0.7). The ionised gas velocity field is dominated by rotation with its major-axis aligned with the stellar kinematic one. The \Hb\ velocity dispersion map shows a depression in the star formation ring (50~km$\,$s$^{-1}$), and slightly rises towards the very centre (80~km$\,$s$^{-1}$, Fig.~\ref{kinHb2}). The \OIII\ velocity dispersion map shows not specific features, with roughly constant values around 50~km$\,$s$^{-1}$.

\subsection{NGC\,5806}
This non-active galaxy is the control for NGC\,3627. \cite{5806_1} observed this galaxy with the HST/NICMOS2 camera in the near-infrared. They detect a prominent bulge and a stellar bar structure in the circumnuclear regions. They classified NGC\,5806 as a galaxy with a concentrated nuclear star formation mixed with dust. Our \Sauron\  data is presented in Figs.~\ref{pair2}, \ref{kinHb2} and  . 
The stellar morphology and kinematics are quite regular. The stellar continuum within the \Sauron\  field of view is elongated along the outer photometric major-axis of the galaxy (PA~$=170\degr$). The stellar velocity field shows a regular rotation pattern with the kinematic-major axis aligned with the photometric major-axis. The velocity dispersion is rising toward the nucleus reaching about 120~km$\,$s$^{-1}$ in the centre. 

The \OIII\ and \Hb\ emission line distribution show elongated features along the photometric PA. That feature exhibits a ring-like structure in \Hb, broken in the North-East part, at a radius $\sim 3\arcsec$. The \OIII/\Hb\ ratio shows rather low values in the \Hb ring ($\sim 0.3$), consistent with star formation. The gas velocity maps are dominated by rotation, and resemble the stellar velocity field, though with a higher amplitude. The ionised gas velocity dispersion maps show no particular structures.

\subsection{NGC\,6951}
NGC\,6951 is a spiral galaxy which has a large stellar bar (PA~$\sim 90\degr$). Its nucleus is classified as Seyfert~2 \citep{Ho3}, but \cite{6951_2} suggest that it could be considered as a transition object, between a very high excitation LINER and a Seyfert 2. The ionised gas (\Ha) is mainly concentrated in the nuclear region and form an annulus of radius $\sim 5\arcsec$, in which massive star formation is taking place. \cite{6951_2} interpreted the \Ha\ kinematics in the circumnuclear zone as a series of nested disks, decoupled from one another. 

The stellar continuum presents isophotes elongated along a PA consistent with the galactic disk major-axis \citet[PA$\sim 138\degr$][]{6951_1}. The stellar velocity field shows regular rotation pattern whit the major-axis aligned with the line-of-nodes of the galaxy. The stellar velocity dispersion shows a drop in the inner 5\arcsec.  

The \Hb\ emission line distribution map shows a inner broken ring elongated along the major-axis of the galaxy ($PA = 137\degr$) while \OIII\ is dominated by the AGN contribution, its intensity maps showing a compact central peak in the inner 3\arcsec, with no particular feature outside. The emission line ratio \OIII/\Hb\ reaches very low values in the ring (0.2), consistent with the presence of on-going star formation in the annulus, and rises towards the centre (values of up to 6) where the effect of the AGN is prominent. The velocity fields are quite regular, dominated by the global rotation, and aligned with the stellar kinematics. The emission line velocity dispersion rises rapidly towards the centre reaching about 100~km$\,$s$^{-1}$ close to the nucleus for both the \Hb\ and \OIII\ emission lines. These values are in good agreement with those found by \cite{6951_2}.  

\section{\Hb\ Kinematic Maps}
In this section we present the \Hb\ kinematic maps for each galaxy. 
\label{app:HB_kin}
\begin{subfigures}
\begin{figure*}
\centering
\includegraphics[height=22cm]{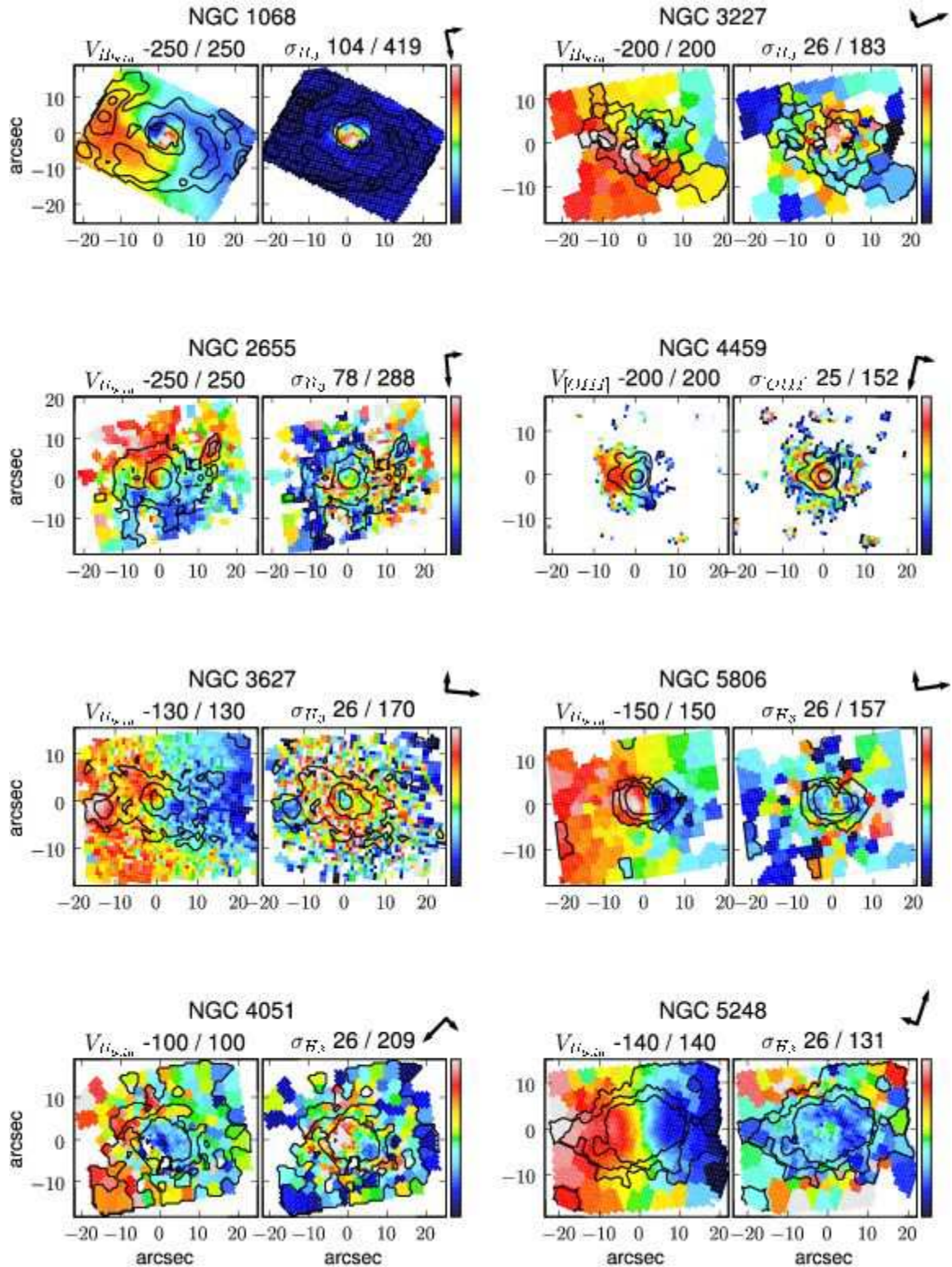}
\caption{Velocity and velocity dispersion of \Hb\ emission lines. Each row present a pair : Seyfert on the left, Control on the right, except for NGC\,1068 and NGC\,3227 which are shown together in the first row. For NGC\,4459, the \OIII\ velocity and velocity dispersion maps are shown (see Sec. \ref{sec:result_intro}.  All the maps are orientated so that the photometric major-axis of the galaxy is on the x-axis. The units are in km\,s$^{-1}$. The color scale is shown on the right hand side and the cut levels are indicated in the top right corner of each map. The long and short arrows on the right of each galaxy name show the North and the East directions, respectively.}
\label{kinHb1}
\end{figure*}
\begin{figure*}
\centering
\includegraphics[height=22cm]{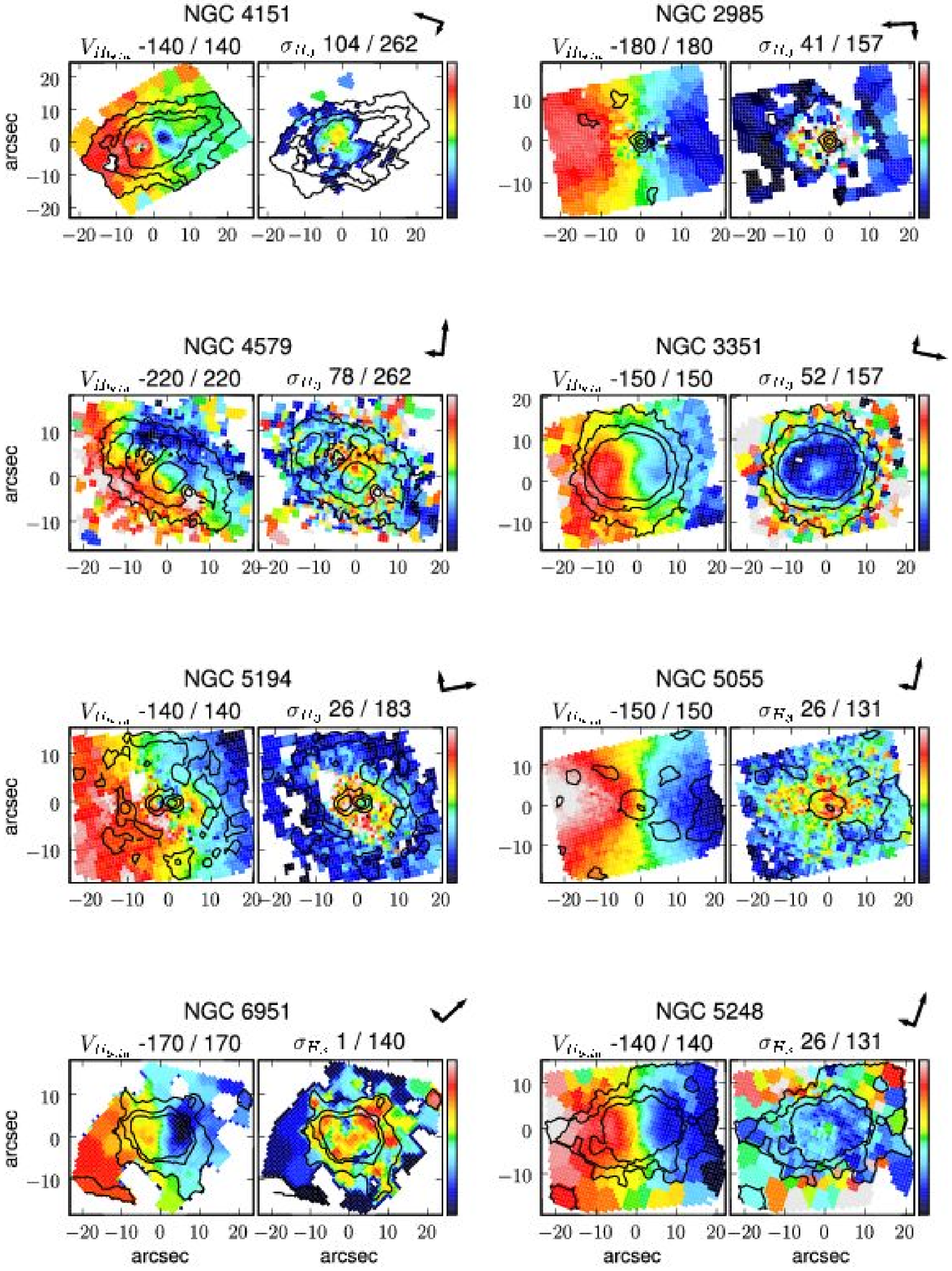}
\caption{Fig. \ref{kinHb1} continued.}

\label{kinHb2}
\end{figure*}
\end{subfigures}